\documentclass[%
 reprint, 
 twocolumn,
 amsmath,amssymb,amsfonts
 aps,
 pre,
floatfix,
showpacs,
showkeys 
]{revtex4-1}
\usepackage{color}
\usepackage{graphicx}
\usepackage{dcolumn}
\usepackage{bm}
\usepackage{hyperref}
\usepackage{tikz}
\usepackage{pifont} 
\usepackage{afterpage}
\usepackage{stmaryrd} 

\begin{document}


\title{Recursive regularization step for high-order lattice Boltzmann methods}

\author{Christophe Coreixas}
\email{coreixas@cerfacs.fr}
\author{Gauthier Wissocq}
\email{wissocq@cerfacs.fr}
\author{Guillaume Puigt}
\email{puigt@cerfacs.fr}
\author{Jean-Fran\c{c}ois Boussuge}
\email{boussuge@cerfacs.fr}
\affiliation{CERFACS, 42 Avenue G. Coriolis, 31057 Toulouse cedex, France.}
\author{Pierre Sagaut}
\email{pierre.sagaut@univ-amu.fr}
\affiliation{Aix-Marseille Univ, CNRS, Centrale Marseille, M2P2 UMR 7340, 13451 Marseille, France.}

\date{\today}

\begin{abstract}
A lattice Boltzmann method (LBM) with enhanced stability and accuracy is presented for various Hermite tensor-based lattice structures. The collision operator relies on a regularization step, which is here improved through a recursive computation of non-equilibrium Hermite polynomial coefficients. In addition to the reduced computational cost of this procedure with respect to the standard one, the recursive step allows to considerably enhance the stability and accuracy of the numerical scheme by properly filtering out second- (and higher-) order non-hydrodynamic contributions in under-resolved conditions. This is first shown in the isothermal case where the simulation of the doubly periodic shear layer is performed with a Reynolds number ranging from $10^4$ to $10^6$, and where a thorough analysis of the case at $Re=3\times 10^4$ is conducted. In the latter, results obtained using both regularization steps are compared against the BGK-LBM for standard (D2Q9) and high-order (D2V17 and D2V37) lattice structures, confirming the tremendous increase of stability range of the proposed approach. Further comparisons on thermal and fully compressible flows, using the general extension of this procedure, are then conducted through the numerical simulation of Sod shock tubes with the D2V37 lattice. They confirm the stability increase induced by the recursive approach as compared with the standard one. 
\end{abstract}
\maketitle


\section{\label{sec:intro}INTRODUCTION}

Over the past two decades, the Lattice Boltzmann Method (LBM) has emerged as an interesting candidate for Computational Fluid Dynamics (CFD) and beyond. 
Despite a first restriction to isothermal and weakly compressible flows, its range of applicability in both physics and engineering has grown in such a way 
that it is now possible to simulate very complex phenomena including turbulence~\cite{YU_PRE_71_2005,SAGAUT_CMA_59_2010}, 
combustion~\cite{PONCEDAWSON_JCP_98_1993,FILIPPOVA_JCP_158_2000,LIN_CF_164_2016}, multiphase 
interactions~\cite{SHAN_PRE_47_1993,SHAN_PRE_49_1994,CHEN_IJHMT_76_2014,MAZLOOMI_PRE_92_2015}, hemodynamics~\cite{YE_JB_49_2016}, 
magnetohydrodynamics~\cite{DELLAR_JCP_179_2002,DELLAR_JSP_121_2005,DELLAR_JSM_2009,DELLAR_EPL_90_2010}, 
relativistic flows~\cite{SUCCI_EPJST_223_2014,MOHSENI_PRE_92_2015} and even quantum systems~\cite{SUCCI_CPC_146_2002,LAPITSKI_TRS_369_2011}.\\
From the numerical point of view, the LBM requires a strong coupling between discretizations of the velocity and the physical spaces. This is usually done using a cartesian grid coupled with an octree-based refinement technique~\cite{LAGRAVA_PhD_2012}. Combining these numerical tools with the particulate nature and the local dynamics of the LBM, the simulation of complex phenomena around realistic geometries is greatly eased~\cite{SENGISSEN_AIAA_2993_2015,CASALINO_AIAA_3101_2014,RIBEIRO_AIAA_1438_2017}.
Moreover, the method is simple to implement, induces a very low computational cost per degree of freedom, and presents a compact stencil, all of which contribute to its intrinsic advantage for 
parallel computations~\cite{SCHORNBAUM_SIAM_38_2016}. All these key points make the LBM of great interest for both academic and industry groups.

The LBM derives from the Boltzmann Equation (BE), the milestone of the kinetic theory of gases~\cite{BOLTZMANN_1872}. The BE describes the balance 
between the transport and the collision of packets of particles through the evolution of the Velocity Distribution Function (VDF). The latter, 
usually written $f(\boldsymbol{x},\boldsymbol\xi,t)$, can be seen as the probability density of finding \emph{fictive} particles at a given point $(\boldsymbol{x},t)$ and with a given speed $\boldsymbol\xi$. Hydrodynamic variables (such as density $\rho$, momentum $\rho\boldsymbol{u}$ and total energy $\rho E$) are then recovered through the computation of the average, over the velocity space, of their mesoscopic counterparts. The Lattice Boltzmann Equation (LBE) results from a velocity discretization of the BE, 
meaning that the degrees of freedom allowed to the transport phenomenon are restricted to a finite set of velocities (directions and norms are fixed).

In order to recover the proper macroscopic set of equations from the LBE, the lattice of discrete speeds $\boldsymbol\xi_i$ and their associated equilibrium 
states $f^{eq}_i$ must be chosen accordingly. To do so, three main approaches are proposed in the literature.\\
\noindent The first suggestion is based on the expansion of the equilibrium state $f^{eq}_i$ into a polynomial, which is performed through a Taylor expansion, assuming that the Mach number is small. The polynomial coefficients are defined \emph{a posteriori} by requiring that the continuum limit of the LBM recovers the Navier-Stokes-Fourier 
set of equations using a predefined lattice of discrete velocities~\cite{QIAN_EPL_17_1992}. This way of building a LBM does not allow to easily link 
lattices with their associated equilibrium states in a \emph{systematic way}, especially when the complexity of the macroscopic behavior of interest increases.\\ 
The second approach is called the Entropic LBM (ELBM) and relies on ensuring the validity of the H-theorem using a discretized velocity space. This is done through a careful evaluation of both the equilibrium state $f^{eq}_i$ and the relaxation time $\tau$~\cite{KARLIN_PRE_88_2013,FRAPOLLI_PRE_90_2014,FRAPOLLI_PRE_93_2016}, leading to an unconditionally stable LBM. Despite a broad range of applicability, the ELBM needs to solve a minimization problem for any grid point and at any time step, hence highly increasing the computational cost per degree of freedom. Such a drawback was recently overcome through the use of an approximated analytic solution of the minimization 
problem~\cite{KARLIN_PRE_90_2014a}, but there seems to be some theoretical work remaining to properly define the validity range of this approximation~\cite{MATTILA_PRE_91_2015}.\\
Finally, the last idea originates from Grad's work~\cite{GRADb_CPAM_2_1949} and was reintroduced into the LBM framework more recently~\cite{SHAN_PRL_80_1998,SHAN_JFM_550_2006,PHILIPPI_PRE_73_2006}. It relies on the projection of the BE onto the Hilbert space spanned by Hermite polynomials. Such a procedure 
allows to build a \emph{systematic} link between the kinetic theory, fluid mechanics and beyond. Indeed, once the macroscopic behavior of interest is chosen, the associated Hermite polynomial basis is then fixed. Typically, a second-order approximation is mandatory for weakly compressible and isothermal flows, a third-order one for isothermal flows, a fourth-order for thermal and fully compressible flows (Navier-Stokes-Fourier) and so on. The final step is to build the associated lattice of speeds by solving a linear system of equations. This system is constrained in order to keep the orthogonality properties of the previously chosen Hermite polynomial basis. This way of proceeding allows to straightforwardly build high-order LBMs~\cite{PHILIPPI_PRE_73_2006}.

The present work is based on this Hermite basis framework and focuses on the regularization of the pre-collision VDFs, which consists in filtering out non-hydrodynamic contributions of the streaming step to stabilize the LBM. Originally, this Projection-based Regularization (PR) procedure was designed for the simulation of particulate suspension~\cite{LADD_JSP_104_2001}. In fact, the stabilization property of the PR was only recently understood. Latt \& Chopard~\cite{LATT_ARXIV_2005,LATT_MCS_72_2006} emphasized its importance for the simulation of flows at high Reynolds number, while Chen and coworkers~\cite{CHEN_PA_362_2006} focused on flows at high Knudsen number. It is now one of the standard stabilization procedures available for both academic studies~\cite{ZHANG_PRE_74_2006,NIU_PRE_76_2007,MONTESSORI_PRE_92_2015} and engineering computations~\cite{SENGISSEN_AIAA_2993_2015,CASALINO_AIAA_3101_2014,RIBEIRO_AIAA_1438_2017}. Even more recently, a Recursive Regularized (RR) collision operator was introduced by Malaspinas~\cite{MALASPINAS_ARXIV_2015} for isothermal LBMs only. The main improvement provided by the RR approach concerned its spectral properties and thus its enhanced stability range, regarding both Reynolds and Mach numbers, for a relatively low additional computational cost. Basically, the RR and PR processes differ in the way off-equilibrium Hermite coefficients (denoted $\boldsymbol{a}_1$ hereafter) are computed.
 
This paper studies the impact of non-equilibrium Hermite coefficients on the stability range of standard and high-order LBMs.
Furthermore, a thorough mathematical derivation of the RR procedure extension to thermal and fully compressible flows is provided and numerically validated. The rest of the paper is organized as follows. In Sec.~\ref{sec:theo}, the recovery of the Navier-Stokes set of equations through the projection of the BE onto the Hermite polynomial basis and the concept behind both regularization steps are reminded. Sec.~\ref{sec:NumericalAssessment} is dedicated to numerics, and starts with the space/time discretization of the LBE (Sec.~\ref{sec:NumericalDiscretization}). In Sec.~\ref{sec:IsothermalValidation}, the discrepancies between the stability range of the PR and the RR versions of the regularized collision operator are highlighted for the simulation of \emph{isothermal} flows using standard and high-order lattices. In Sec.~\ref{sec:FCompressibleValidation}, the extension of the RR step to \emph{thermal and fully compressible} high-order LBMs is presented and illustrated on two configurations of the Sod shock tube. For the sake of completeness, appendixes on the mathematical proof of the Hermite coefficients recursive formulas (App.~\ref{sec:Recursivity_a1_proof} \&~\ref{sec:appendixB}), implementation details (App.~\ref{sec:appendixC}) and Hermite tensor basis for standard and high-order lattice structures (App.~\ref{sec:appendixD}) are finally provided.  

\section{\label{sec:theo}THEORETICAL BACKGROUND}

In this paper, all quantities are defined in the $D$-dimensional Cartesian space $\mathbb{R}^D$ 
and by definition, for any vector $\boldsymbol{v}$, $v^2 = \boldsymbol{v} \cdot \boldsymbol{v}$ is the square of the norm of $\boldsymbol{v}$. Properties concerning tensor products and Hermite polynomials in $\mathbb{R}^D$
are based on Grad's and Shan's works~\cite{GRADa_CPAM_2_1949, SHAN_JFM_550_2006}.

\subsection{\label{sec:Boltzmann}Boltzmann equation}

In kinetic theory,  gases are modeled by the VDF $f(\boldsymbol{x},\boldsymbol\xi,t)$ describing the probability density of finding a \emph{fictive} particle at position $\boldsymbol{x}$, time $t$ and with a mesoscopic velocity $\boldsymbol{\xi}$. When no external accelerations are considered, this VDF evolves through time and space in accordance to the force-free form of the BE:
\begin{equation}
\partial_t f + \boldsymbol\xi\cdot\boldsymbol\nabla f = \Omega_f, 
\label{eq:BE}
 \end{equation}
where $\cdot$ denotes the scalar product over  $\mathbb{R}^D$, $\boldsymbol\nabla$  is the gradient operator associated to the physical space and $\Omega_f$ is the collision operator. The macroscopic quantities of interest (density $\rho$, momentum $\rho\boldsymbol{u}$, and total energy $\rho E$) are recovered averaging their mesoscopic counterparts over the velocity space:
\begin{equation}
\left\{
\begin{array}{r @{\: = \:} l}
\rho & \displaystyle{\int
f \,\mathrm{d}\boldsymbol\xi}, \vspace{0.1cm}\\
\rho\boldsymbol{u} & \displaystyle{\int
f \boldsymbol{\xi}\,\mathrm{d}\boldsymbol\xi}, \vspace{0.1cm}\\
2 \rho E & \displaystyle{\int
 f \xi^2\,\mathrm{d}\boldsymbol\xi}, 
\end{array}
\right.
\end{equation}
with integrals computed over $\mathbb{R}^D$. Hereafter, integration bounds will be omitted for the sake of clarity. 

Regarding the collision model $\Omega_f$, it must satisfy the conservation of mass, momentum and total energy:
\begin{equation}
\displaystyle{\int
\Omega_f\,\Phi(\boldsymbol\xi)\,\mathrm{d}\boldsymbol\xi} = \mathbf{0},
\label{eq:CollisionInvariant}
\end{equation}
with $\Phi(\boldsymbol\xi)=(1,\boldsymbol\xi,\xi^2/2)$. This collision process induces a relaxation of the VDF to the local thermodynamic equilibrium
\begin{equation}
f^{(eq)}=\frac{\rho}{(2 \pi r T)^{D/2}} \exp \left( - \frac{c^2}{2rT} \right),
\label{eq:feq}
\end{equation}\\
where $\boldsymbol{c}=\boldsymbol{\xi}-\boldsymbol{u}$, $r$ is the gas constant and $T$ the thermodynamic temperature. 
Throughout this paper, only linearized collision operators, such as the single relaxation time collision term of Bhatnagar-Gross-Krook 
(BGK)~\cite{BHATNAGAR_PR_94_1954} 
\begin{equation}
\Omega_f^{BGK}=-\frac{1}{\tau}\left( f-f^{(eq)} \right),
\label{eq:singleBGK}
\end{equation}
will be considered.

\subsection{\label{sec:ProjHermite}Projection onto the Hermite polynomial basis}

\indent In the present context, solutions of Eq.~(\ref{eq:BE}) are sought in the form of Hermite polynomials~\cite{SHAN_PRL_80_1998},
\begin{equation}
 f(\boldsymbol{x}, \boldsymbol\xi,t) = \omega(\boldsymbol\xi)\displaystyle{\sum_{n=0}^{\infty} \dfrac{1}{n!(rT_0)^n}\boldsymbol{a}^{(n)}(\boldsymbol{x},t):\boldsymbol{\mathcal{H}}^{(n)}(\boldsymbol\xi)},
 \label{eq:HermiteDevelopment}
 \end{equation}
where $:$ stands for the full contraction of indexes, $T_0$ is a reference temperature, $\boldsymbol{a}^{(n)}$ is the (tensor of) coefficient(s) related to the Hermite tensor $\boldsymbol{\mathcal{H}}^{(n)}$, both being $n-$rank tensors, and $\omega(\boldsymbol\xi)$ is the weight function. They are defined as follows:
\begin{equation}
\boldsymbol{a}^{(n)}(\boldsymbol{x},t) =  \displaystyle{\int  f(\boldsymbol{x}, \boldsymbol\xi,t) \boldsymbol{\mathcal{H}}^{(n)}(\boldsymbol\xi) \,\mathrm{d}\boldsymbol\xi},
\label{eq:HermiteCoefficientDef}
\end{equation}
where
\begin{equation}
\boldsymbol{\mathcal{H}}^{(n)} = \dfrac{(-rT_0)^n}{\omega(\boldsymbol\xi)}\boldsymbol\nabla^n_{\boldsymbol\xi}\omega(\boldsymbol\xi)
\label{eq:RodriguesFormula}
\end{equation}
with
\begin{equation}
\omega(\boldsymbol\xi) = \dfrac{1}{(2\pi r T_0)^{D/2}}\exp \left( -\frac{\xi^2}{2rT_0} \right),
\end{equation}
$\boldsymbol\nabla^n_{\boldsymbol\xi}$ being the $n$-th derivative with respect to the velocity space.
Eq.~(\ref{eq:HermiteDevelopment}) can be seen as the decomposition of $f$ onto an orthogonal polynomial basis, since Hermite tensors are orthogonal 
with respect to the following scalar product:
\begin{equation}
\langle g \vert h \rangle \equiv \displaystyle{\int \omega(\boldsymbol\xi)g(\boldsymbol\xi)h(\boldsymbol{\xi})\mathrm{d}\boldsymbol\xi}.
\label{eq:ScalarProduct}
\end{equation} 
Thus $\boldsymbol{a}^{(n)}$ can simply be obtained as a projection of $f$ onto this orthogonal basis $\boldsymbol{a}^{(n)} = \langle \boldsymbol{\mathcal{H}}^{(n)} \vert f/ \omega \rangle$.
For the sake of clarity, the list of function variables will be omitted throughout the rest of the paper, except in Sec~\ref{sec:NumericalDiscretization} where the space/time discretization of the LBE is presented.

By construction, $\boldsymbol{a}^{(n)}$ can be linked to the familiar hydrodynamic moments~\cite{SHAN_JFM_550_2006,MALASPINAS_PhD_2009}:
\begin{equation}
\centering
\left\{
\begin{array}{r @{\: = \:} l}
\boldsymbol{a}^{(0)} & \rho, \vspace{0.1cm}\\
\boldsymbol{a}^{(1)} & \rho\boldsymbol{u}, \vspace{0.1cm}\\
\boldsymbol{a}^{(2)} & \boldsymbol{\Pi} + \rho(\boldsymbol{u}^2-\boldsymbol{\delta}), \vspace{0.1cm}\\
\boldsymbol{a}^{(3)} & \boldsymbol{Q} + \boldsymbol{ua}^{(2)} + (1-D)\rho\boldsymbol{u}^3, \vspace{0.1cm}\\
\boldsymbol{a}^{(4)} & \boldsymbol{R} - \boldsymbol{P\delta} + \boldsymbol{\delta}^2,
\end{array}
\right.
\end{equation}
where 
$\boldsymbol{\delta}$ is the identity matrix, $\boldsymbol{\delta}^2$ is the fourth-order identity tensor, and
\begin{equation}
\boldsymbol{\Pi} = \displaystyle{\int f \boldsymbol{c}^2\,\mathrm{d}\boldsymbol\xi},\quad 
\boldsymbol{Q} = \displaystyle{\int f \boldsymbol{c}^3\,\mathrm{d}\boldsymbol\xi},\quad
\boldsymbol{R} = \displaystyle{\int f \boldsymbol{c}^4\,\mathrm{d}\boldsymbol\xi}.
\end{equation}


In order to create a \emph{systematic} link between the BE and its macroscopic counterpart, Eq.~(\ref{eq:BE}) is projected onto the Hermite tensor basis using the projection operator $\langle \cdot \vert \cdot \rangle$ defined in Eq.~(\ref{eq:ScalarProduct}):
\begin{equation}
 \partial_t \left(\boldsymbol{a}^{(n)}\right) + \boldsymbol\nabla\cdot \left(\boldsymbol{a}^{(n+1)}\right) + rT_0\boldsymbol\nabla\boldsymbol{a}^{(n-1)} =\Omega^{BGK}.
 \label{eq:HermiteBE}
\end{equation}
Here the BGK approximation~(\ref{eq:singleBGK}) is adopted for the computation of the collision term, \emph{i.e.}, $\Omega^{BGK} = -( \boldsymbol{a}^{(n)} -  \boldsymbol{a}_{eq}^{(n)})/\tau$, where $\boldsymbol{a}_{eq}^{(n)} = \langle \boldsymbol{\mathcal{H}}^{(n)} \vert f^{(eq)}/\omega \rangle$. The computation of these coefficients is straightforward since the expression of $f^{(eq)}$ is known. Up to the fourth order, one obtains
\begin{equation}
\centering
\left\{
\begin{array}{r @{\: = \:} l}
\boldsymbol{a}_{eq}^{(0)} & \rho, \vspace{0.1cm}\\
\boldsymbol{a}_{eq}^{(1)} & \rho\boldsymbol{u}, \vspace{0.1cm}\\
\boldsymbol{a}_{eq}^{(2)} & \rho\big[\boldsymbol{u}^2+rT_0\left(\theta-1\right)\boldsymbol{\delta}\big], \vspace{0.1cm}\\
\boldsymbol{a}_{eq}^{(3)} & \rho\big[\boldsymbol{u}^3+rT_0\left(\theta-1\right)\boldsymbol{u\delta}\big], \vspace{0.1cm}\\
\boldsymbol{a}_{eq}^{(4)} & \rho\big[\boldsymbol{u}^4+rT_0\left(\theta-1\right)\boldsymbol{u}^2\boldsymbol{\delta}+(rT_0)^2\left(\theta-1\right)^2\boldsymbol{\delta}^2\big],
\end{array}
\right.
\label{eq:HermiteCoeffEq}
\end{equation}
with $\theta = T/T_0$, $T_0$ being a reference temperature.

\subsection{\label{sec:CE}Chapman-Enskog expansion}
\indent After expressing the BE in the Hermite tensor basis and defining the related 
coefficients $\boldsymbol{a}^{(n)}$, the macroscopic behavior linked to Eq.~(\ref{eq:HermiteBE}) is recovered by applying a separation of scales through the 
Chapman-Enskog expansion~\cite{CHAPMAN_Book_3rd_1970}.
For this purpose, the time derivative is expanded in powers of the Knudsen number 
$\epsilon$ as in~\cite{CHAPMAN_Book_3rd_1970}:
\begin{equation}
\partial_{t} = \epsilon \partial_{t_1} + \epsilon^2 \partial_{t_2}, \quad \boldsymbol\nabla = \epsilon\boldsymbol\nabla_1.
\label{eq:CE} 
\end{equation}

First, let us assume that $f$ and $\boldsymbol{a}^{(n)}$ are at equilibrium. Injecting Eq.~(\ref{eq:CE}) into Eq.~(\ref{eq:HermiteBE}), this separation of scales leads to\\\\
\underline{$\epsilon^{0}$}:
\begin{equation}
\boldsymbol{a}_{0}^{(n)} =  \boldsymbol{a}_{eq}^{(n)},
\end{equation}
\underline{$\epsilon^{1}$}:
\begin{equation}
\partial_{t_1} \boldsymbol{a}_{0}^{(n)} + \boldsymbol\nabla_1\cdot \left(\boldsymbol{a}_{0}^{(n+1)}\right) + rT_0\boldsymbol\nabla_1\boldsymbol{a}_{0}^{(n-1)} =0,
\label{eq:HermiteBE_Euler1}
\end{equation}
\underline{$\epsilon^{2}$}:
\begin{equation}
\partial_{t_2} \boldsymbol{a}_{0}^{(n)}  =0.
\label{eq:HermiteBE_Euler2}
\end{equation}
\noindent Adding all contributions, the macroscopic equations corresponding to the conservation of mass, momentum and total energy are finally recovered, using Eq.~(\ref{eq:HermiteCoeffEq}) for $n=0$, $n=1$ and $n=2$ respectively,
\begin{equation}
\begin{array}{c}
\partial_t{}(\rho) + \boldsymbol{\nabla}\cdot(\rho\boldsymbol{u})=0, \vspace{0.1cm}\\
\partial_t(\rho\boldsymbol{u}) + \boldsymbol{\nabla}\cdot(\rho\boldsymbol{u}^2 + p\boldsymbol{\delta})= 0, \vspace{0.1cm}\\
\partial_t(\rho E) + \boldsymbol{\nabla}\cdot[(\rho E+ p)\boldsymbol{u}]=0,
\end{array}
\label{eq:HermiteBE_Euler}
\end{equation}
where $p=\rho r T$ is the thermodynamic pressure. Hence, assuming $f$ and $\boldsymbol{a}^{(n)}$ are at equilibrium leads to Euler's set of equations. 

In a second step, the contributions of order $\mathcal{O}(\epsilon)$ are taken into account in the definition of $f$ and $\boldsymbol{a}^{(n)}$
\begin{equation}
\begin{array}{l @{,\:} r}
f = f^{(0)} + f^{(1)} & f^{(0)} \gg f^{(1)}\sim\mathcal{O}(\epsilon), \vspace{0.1cm}\\
\boldsymbol{a}^{(n)} =  \boldsymbol{a}_{0}^{(n)} + \boldsymbol{a}_{1}^{(n)} & \boldsymbol{a}_{0}^{(n)} \gg \boldsymbol{a}_{1}^{(n)}\sim \mathcal{O}(\epsilon),
\end{array} 
\label{eq:CE_NS}
\end{equation}
assuming the continuum limit $\epsilon \ll 1$. Applying the same multiscale analysis leads to\\\\
\underline{$\epsilon^{1}$}:
\begin{equation}
\partial_{t_1} \boldsymbol{a}_{0}^{(n)} + \boldsymbol\nabla_1\cdot \left(\boldsymbol{a}_{0}^{(n+1)}\right) + rT_0\boldsymbol\nabla_1\boldsymbol{a}_{0}^{(n-1)} =- \dfrac{\boldsymbol{a}_{1}^{(n)}}{\tau},
\label{eq:HermiteBE_NS1}
\end{equation}
\underline{$\epsilon^{2}$}:
\begin{equation}
\partial_{t_2} \boldsymbol{a}_{0}^{(n)} + \partial_{t_1} \boldsymbol{a}_{1}^{(n)} + \boldsymbol\nabla_1\cdot \left(\boldsymbol{a}_{1}^{(n+1)}\right) + rT_0\boldsymbol\nabla_1\boldsymbol{a}_{1}^{(n-1)} =0.
\label{eq:HermiteBE_NS2}
\end{equation}
Therefore the set of macroscopic equations~(\ref{eq:HermiteBE_Euler}) becomes 
\begin{equation}
\begin{array}{c}
\partial_t(\rho) + \boldsymbol{\nabla}\cdot(\rho\boldsymbol{u})=0, \vspace{0.1cm}\\
\partial_t(\rho\boldsymbol{u}) + \boldsymbol{\nabla}\cdot(\rho\boldsymbol{u}^2 + p\boldsymbol{\delta})=  - \boldsymbol{\nabla}\cdot\boldsymbol{a}_{1}^{(2)} , \vspace{0.1cm}\\
\partial_t(\rho E) + \boldsymbol{\nabla}\cdot[(\rho E+ p)\boldsymbol{u}]=  - \dfrac{1}{2} \mathrm{Tr}\left(\boldsymbol{\nabla}\cdot\boldsymbol{a}_{1}^{(3)}\right),
\end{array}
\label{eq:CE_NS}
\end{equation}
since the collision model must satisfy Eq.~(\ref{eq:CollisionInvariant}) or equivalently,
\begin{equation}
\bm{a}_1^{(0)} = \bm{a}_{1}^{(1)} = \mathrm{Tr}\big(\bm{a}_{1}^{(2)}\big) = \bm{0}.
\label{eq:NullityOfa1}
\end{equation}
Now, only $a_{1,\alpha\beta}^{(2)}$ and $a_{1,\alpha\beta\beta}^{(3)}$ remain to be computed. They can either be computed thanks to Eq.~(\ref{eq:HermiteBE}) or noticing that,
\begin{equation}
\begin{array}{l @{\: = \:} l @{\: = \:} l @{\: \equiv \:} l}
\boldsymbol{\Pi}^{(1)} & \displaystyle{\int f^{(1)}\boldsymbol{c}^2\mathrm{d}\boldsymbol\xi} & \displaystyle{\int f^{(1)}\boldsymbol{\xi}^2\mathrm{d}\boldsymbol\xi} + 0 & \boldsymbol{a}_{1}^{(2)},\vspace{0.1cm}\\
\boldsymbol{Q}^{(1)} & \displaystyle{\int f^{(1)}\boldsymbol{c}^3\mathrm{d}\boldsymbol\xi} & \displaystyle{\int f^{(1)}\boldsymbol{\xi}^3\mathrm{d}\boldsymbol\xi} + 0 & \boldsymbol{a}_{1}^{(3)},
\end{array}
\end{equation}
where $\boldsymbol{\Pi}^{(1)}$ and $\boldsymbol{q}^{(1)}=\mathrm{Tr}(\mathbf{Q}^{(1)})/2$ are the standard second- and third-order off-equilibrium moments at the Navier-Stokes level~\cite{MALASPINAS_PhD_2009},
\begin{equation}
\left\{
\begin{array}{r @{\: = \:} l}
\boldsymbol{\Pi}^{(1)} & -\tau p\:\left[\boldsymbol{S}  - \left(\frac{2}{D}\boldsymbol{\nabla}\cdot\boldsymbol{u}\right)\boldsymbol{\delta}\right], \vspace{0.1cm}\\
\boldsymbol{q}^{(1)} & -\tau p c_p\boldsymbol{\nabla}T + \boldsymbol{u}\cdot\boldsymbol{\Pi}^{(1)},
\end{array}
\right.
\label{eq:NeqMomentsNS}
\end{equation}
with $\boldsymbol{S} = \boldsymbol{\nabla u}+(\boldsymbol{\nabla u})^{T}$, $\cdot^{T}$ standing for the transpose operator. $c_p=(1+D/2)r$ is the heat capacity at constant pressure.\\
Finally, injecting Eq.~(\ref{eq:NeqMomentsNS}) in Eq.~(\ref{eq:CE_NS}) allows to link the Hermite formulation of the BE~(\ref{eq:HermiteBE}) to the following macroscopic set of equations, namely, Navier-Stokes-Fourier equations,
\begin{equation}
\begin{array}{c}
\partial_t(\rho) + \boldsymbol{\nabla}\cdot(\rho\boldsymbol{u})=0, \vspace{0.1cm}\\
\partial_t(\rho\boldsymbol{u}) + \boldsymbol{\nabla}\cdot(\rho\boldsymbol{u}^2)= \boldsymbol{\nabla}\cdot(\boldsymbol{\sigma}), \vspace{0.1cm}\\
\partial_t(\rho E) + \boldsymbol{\nabla}\cdot[\rho E\boldsymbol{u}]= \boldsymbol{\nabla}\cdot(\lambda'\boldsymbol{\nabla}T)+  \boldsymbol{\nabla}\cdot(\boldsymbol{\sigma}\cdot\boldsymbol{u}),
\end{array}
\label{Eq:NavierStokes}
\end{equation}
where $\boldsymbol{\sigma} = \overline{\boldsymbol{\Pi}}- p\boldsymbol{\delta}$ is the stress tensor, $ \overline{\boldsymbol{\Pi}} =\mu \left[\boldsymbol{S}  - \left(\frac{2}{D}\boldsymbol{\nabla}\cdot\boldsymbol{u}\right)\boldsymbol{\delta}\right]$ is the traceless
viscous stress tensor, $\mu = \tau p$ and $\lambda' = \tau p c_p$ being the dynamic viscosity and the thermal conductivity coefficients.

Before proceeding any further, several remarks can be pointed out.\\ 
First, only coefficients $\boldsymbol{a}_{0}^{(n)}$ ($\boldsymbol{a}_{1}^{(n)}$) up to the fourth order (third order) are needed to recover the Navier-Stokes set of equations. And more generally, the Chapman-Enskog expansion of the BE at order $k$ needs coefficients of the Hermite expansion to include terms at order $n+k$. This is 
mandatory to properly recover the hydrodynamic behavior of the BE~\cite{SHAN_JFM_550_2006}.\\ 
Second, a strong assumption is made by using the BGK collision operator, which allows only one parameter to represent the physical behavior of the fluid: the single relaxation time $\tau$. This choice induces a coupling between the momentum and the energy relaxation processes since the Prandtl number is fixed at $Pr = \mu c_p/\lambda' = 1$. To overcome this deficiency, a more sophisticated collision operator may be employed, such as the general Multi-Relaxation Time (MRT) collision term expressed in the Hermite tensor basis~\cite{SHAN_IJMPC_18_2007}. The latter reads
\begin{equation}
\Omega^{MRT} =  -\displaystyle{\sum_{n=0}^{\infty}} \dfrac{1}{\tau_n} \dfrac{1}{n!(rT_0)^n}\boldsymbol{a}^{(n)}:\boldsymbol{\mathcal{H}}^{(n)}.
\end{equation}
Here choosing $\tau_2 = \mu/p$ and $\tau_3 = \lambda/pc_p$ allows to tune the Prandtl number which now equals $Pr =\tau_2/\tau_3$.
A double distribution function LBM could also be used to overcome this difficulty (see for example~\cite{GUO_PRE_75_2007}).\\ 
Third, as depicted in Eq.~(\ref{Eq:NavierStokes}), the computed viscous stress tensor is traceless, which means that only fluids without bulk viscosity can be simulated. This is another consequence of the BGK approximation, and again this can be avoided by using a more sophisticated collision operator belonging to the Hermite polynomial expansion framework~\cite{KRÜGER_Book_2017}.\\ 
Last, the specific heat ratio has a fixed value $\gamma = (D+2)/D$. To overcome this issue, one can employ a second distribution function to take into account the energy evolution linked to internal degrees of freedom (rotational and vibrational) of molecules~\cite{RYKOV_FD_10_1975,NIE_PRE_77_2008,FRAPOLLI_PRE_93_2016}.

In the rest of the paper, only single-VDF-based LBMs coupled with the single-relaxation-time approximation are considered.

\subsection{\label{sec:Truncation}Truncation of the VDF}

As briefly discussed in Sec.~\ref{sec:CE}, a finite expansion of the VDF in Hermite tensors, up to the fourth order, is sufficient to recover the macroscopic behavior of Navier-Stokes' equations from the BE. Here particular attention is paid to error terms arising, at the macroscopic level, from a wrong truncation of the VDF. From now on, let $f^{N}$ denote the truncation of 
$f$, to the order $N$, of the Hermite polynomial development
\begin{equation}
	f^{N} = \omega \sum_{n=0}^N \frac{1}{n!(rT_0)^n}\boldsymbol{a}^{(n)}:\boldsymbol{\mathcal{H}}^{(n)}.
\end{equation}
The key element is to ensure that this truncation still allows the proper conservation of the macroscopic moment of the VDF. Introducing $\boldsymbol{a}^{(M)}$ as the $M$th-order moment of $f$:
\begin{equation}
	\boldsymbol{a}^{(M)} = \int \boldsymbol{\mathcal{H}}^{(M)} f\, \mathrm{d}\boldsymbol{\xi}, 
\end{equation}
orthogonality properties of the Hermite polynomials lead to~\cite{SHAN_JFM_550_2006}
\begin{equation}
	\int \boldsymbol{\mathcal{H}}^{(M)} f\, \mathrm{d}\boldsymbol{\xi} = \int \boldsymbol{\mathcal{H}}^{(M)} f^{N}\, \mathrm{d}\boldsymbol{\xi},\text{ if } N \geq M.
	\label{eq:Truncation}
\end{equation}
This means that $f$ and $f^N$ share the same moments up to the order $M$, if the truncation order $N$ is at least equal to the highest moment $M$ that we need to conserve.

As mentioned in Sec.~\ref{sec:CE}, moments of $f^{(0)}$ up to the fourth order are sufficient to recover the proper macroscopic behavior of the BE. 
Consequently, truncation errors appear at the macroscopic level when the equilibrium VDF is truncated to an order lower than $N=4$.
These error terms can be evaluated quite easily using Eq.~(\ref{eq:HermiteBE}). For $N = 3$, the only term that cannot be used for the computation of $\boldsymbol{a}_{1}^{(3)}$ is $\boldsymbol\nabla\cdot \boldsymbol{a}_0^{(4)}$. Hence, the error term introduced by a third-order truncation modifies the heat flux $\boldsymbol{q} = \mathrm{Tr}(\mathbf{Q})/2$:
\begin{align}\label{eq:errorN3}
\boldsymbol{q}' &= \boldsymbol{q}^{(1)} -\dfrac{\tau}{2} \mathrm{Tr}\left[\boldsymbol\nabla\cdot\boldsymbol{a}_0^{(4)} \right] \nonumber\\[0.1cm]
					   &= \boldsymbol{q}^{(1)} -\dfrac{\tau}{2}  \boldsymbol\nabla\cdot\bigg\{\rho u^2 \boldsymbol{u}^2+ \rho(\theta-1)\left[(4+D)\boldsymbol{u}^2+u^2\boldsymbol{\delta}\right] \nonumber\\[0.1cm]
					   &\quad\quad\quad\quad\quad\quad +\rho(\theta-1)^2(2+D)\boldsymbol{\delta}\bigg\}\nonumber\\[0.1cm]
					   &= \boldsymbol{q} + \mathcal{O}\left(Ma^4,Ma^2\theta,\theta^2\right).  
\end{align}
For $N = 2$, too many error terms are introduced in the energy equation and the concept of temperature should not be used. This is why this truncation is restricted to isothermal, or more precisely athermal LBMs ($\theta = 1$). Furthermore, the computation of $\boldsymbol{a}_{1}^{(2)}$ is also affected by the truncation. As for the heat flux when $N=3$, the deviation also comes from the divergence term $\boldsymbol\nabla\cdot(\boldsymbol{a}_0^{(3)})$ which cannot be used anymore for the computation of $\boldsymbol{a}_{1}^{(2)}$. The viscous stress tensor is thus modified:
 \begin{align} \label{eq:errorN2}
\boldsymbol{\Pi}' &= \boldsymbol{\Pi}^{(1)} -\tau \boldsymbol{\nabla}\cdot \left(\boldsymbol{a}_0^{(3)}\right)\nonumber\\[0.1cm]
					   &= \boldsymbol{\Pi}^{(1)} -\tau\boldsymbol{\nabla}\cdot\left(\rho\boldsymbol{u}^3\right)\nonumber\\[0.1cm]
					   &\approx \boldsymbol{\Pi}^{(1)} + \mathcal{O}(Ma^3).
\end{align}
The famous $\mathcal{O}(Ma^3)$ error term is then recovered from the above procedure, explaining why second-order truncation of the VDF are restricted to the simulation of weakly compressible flows~\cite{QIAN_EPL_17_1992}.

In the most general case, the truncation criterium~(\ref{eq:Truncation}) becomes~\cite{SHAN_JFM_550_2006} 
\begin{equation}
\forall k\in \mathbb{N},\:	\int \boldsymbol{\mathcal{H}}^{(M)} f^{(k)}\, \mathrm{d}\boldsymbol{\xi} = \int \boldsymbol{\mathcal{H}}^{(M)} f^{(k),N}\, \mathrm{d}\boldsymbol{\xi}
 \end{equation}
if $N+k \geq M$. Here, we recover the fact that the Navier-Stokes macroscopic behavior is achieved if $N=4$ at equilibrium ($k=0$), while $N=3$ is sufficient for the non-equilibrium VDF $f^{(1),N}$. From the macroscopic point of view, there is no reason why $f^{(0),N}$ and $f^{(1),N}$ should be truncated at the same order $N$. But it will be shown hereafter (Sec.~\ref{sec:Regularization}) that keeping the same order $N$ for both $f^{(0),N}$ and $f^{(1),N}$ is possible thanks to a recursive computation of these coefficients.   

\subsection{\label{sec:Regularization}Regularized collision operator}
Both truncations $f^{(0),N}$ and $f^{(1),N}$ will now be considered. The reason lies in the definition of the regularization step~\cite{LATT_ARXIV_2005,LATT_MCS_72_2006,CHEN_PA_362_2006}. 
This collision operator is based on the regularization of the non-equilibrium part of the pre-collision distribution functions, and aims at filtering out non-hydrodynamic sources. To do so, $f^N$ is reconstructed before the collision step discarding $\mathcal{O}(\epsilon^k)$ contributions ($k\geq 2$):
\begin{equation}
f^{reg,N} \equiv f^{(0),N} + f^{(1),N},
\end{equation}
with
\begin{equation}
 	f^{(1),N}=\omega \sum_{n=2}^N \frac{1}{n! (rT_0)^n} \boldsymbol{a}_1^{(n)}:\boldsymbol{\mathcal{H}}^{(n)},
\end{equation} 
and where the sum begins at $n=2$ due to mass and momentum conservation~(\ref{eq:NullityOfa1}).
Post-collision VDFs are then defined as follows,
\begin{align}
f^{coll,N} &= f^{(0),N} + \left(1-\dfrac{1}{\tau}\right)f^{(1),N} \label{eq:RegularizedCollision}\\[0.1cm]
                  &= \omega \sum_{n=0}^N \frac{1}{n! (rT_0)^n} \left[\boldsymbol{a}_0^{(n)} + \left(1-\dfrac{1}{\tau}\right)\boldsymbol{a}_1^{(n)}\right]:\boldsymbol{\mathcal{H}}^{(n)}.\nonumber
\end{align}
Here coefficients $\boldsymbol{a}_1^{(n)}$ are the only missing information required to 
reconstruct $f^{(1),N}$. Originally in~\cite{LATT_ARXIV_2005,LATT_MCS_72_2006}, this stabilization technique was used for the simulation of isothermal and weakly compressible flows for which only second-order terms were kept. Off-equilibrium coefficients $\boldsymbol{a}_1^{(2)}$ were then computed projecting the VDFs onto the second-order Hermite polynomials
\begin{equation}
	\boldsymbol{a}_1^{(2)} \approx \displaystyle{\int} \boldsymbol{\mathcal{H}}^{(2)}\left(f^N-f^{(0),N}\right) \mathrm{d} \boldsymbol{\xi},
\end{equation}
assuming that $f^N-f^{(0),N}\approx f^{(1),N}$. But doing so, only contributions belonging to the Hilbert space that were not taken into account are filtered, while some non-hydrodynamic contributions are still hidden in $\boldsymbol{a}_1^{(2)}$. In the rest of the paper, this method will be referred as to the projection-based regularization (PR) process. 

Later, Malaspinas~\cite{MALASPINAS_ARXIV_2015} proposed a \emph{complete} regularization procedure based on \emph{recursive} properties of the off-equilibrium coefficients, allowing an enhancement of accuracy and stability compared to BGK and standard MRT models. 
Regarding the \emph{recursive} property of this stabilization procedure, it comes from the fact that non-equilibrium coefficients $\boldsymbol{a}_1^{(n)}$ are computed using a recursive formula flowing from the Chapman-Enskog expansion. This step provides a proper way to filter out non-hydrodynamic sources.\\
It must be understood that this recursive regularization (RR) approach was introduced in the context of \emph{isothermal and weakly compressible} flows for \emph{standard} LBMs, while in the present paper the recursive formula is extended to the \emph{thermal and fully compressible} case (see App.~\ref{sec:Recursivity_a1_proof} for its derivation) and applied to \emph{high-order} LBMs. The latter reads
\begin{widetext}
\begin{align}
\label{eq:RecursiveRelation_ThermalCase}
\forall n \ge 4,\quad a_{1,{\alpha_1..\alpha_n}}^{(n)}&=u_{\alpha_{n}} a_{1,{\alpha_1..\alpha_{n-1}}}^{(n-1)} + rT_0 (\theta-1) \sum_{l=1}^{n-1} \delta_{\alpha_l \alpha_n} a_{1,{\beta_l}}^{(n-2)} + \frac{1}{\rho} \sum_{l=1}^{n-1} a_{0,{\beta_l}}^{(n-2)} a_{1,{\alpha_l \alpha_n}}^{(2)} \nonumber\\
	&+ \frac{1}{\rho} \sum_{l=1}^{n-1} \sum_{m>l}^{n-1} a_{0,{\beta_{lm}}}^{(n-3)} \left( a_{1,{\alpha_l \alpha_m \alpha_n}}^{(3)} - u_{\alpha_l} a_{1,{\alpha_m \alpha_n}}^{(2)} - u_{\alpha_m} a_{1,{\alpha_l \alpha_n}}^{(2)} - u_{\alpha_n} a_{1,{\alpha_l \alpha_m}}^{(2)} \right),
\end{align}
\end{widetext}
where $\boldsymbol{a}_1^{(2)}$ and $\boldsymbol{a}_1^{(3)}$ can either be computed thanks to the projection of $f^{(1)}$ onto the Hermite polynomial basis or using finite differences. In the isothermal case ($\theta=1$), Malaspinas' recursive relation~\cite{MALASPINAS_ARXIV_2015} is recovered
for $n \ge 3$:
\begin{equation}
\label{eq:MalaspinasRecursiveRelation}
	a_{1,{\alpha_1..\alpha_n}}^{(n)}=u_{\alpha_{n}} a_{1,{\alpha_1..\alpha_{n-1}}}^{(n-1)}  + \frac{1}{\rho} \sum_{l=1}^{n-1} a_{0, \beta_l} ^{(n-2)} a_{1,\alpha_l \alpha_n}^{(2)}. 
\end{equation}
Again, $\boldsymbol{a}_1^{(2)}$ coefficients can be computed by projection or using finite differences.\\ 
Furthermore, a correct evaluation of $\boldsymbol{a}_1^{(2)}$ requires a proper evaluation of $\boldsymbol{a}_0^{(3)}$ since
\begin{equation}
	\boldsymbol{a}_1^{(2)} = - \tau \left[ \partial_t \boldsymbol{a}_0^{(2)} + \boldsymbol{\nabla} \cdot \boldsymbol{a}_0^{(3)} \right].
\end{equation}
Thus, in the isothermal case, the equilibrium VDF should theoretically be developed up to the third order. Similarly, in the thermal case, correct calculations of $\boldsymbol{a}_1^{(2)}$ and $\boldsymbol{a}_1^{(3)}$ require a development of $f^{(0)}$ up to the fourth order, which is no more binding than the condition to recover the thermal and fully compressible Navier-Stokes-Fourier equations. 

\subsection{\label{sec:DVBE}Discretization of the velocity space}
In order to numerically solve the BE, a discretization of the velocity space is necessary. It consists in keeping only a discrete set of $V$ velocities $\boldsymbol{\xi}_i,\ i \in \llbracket 1,V \rrbracket$, ensuring the preservation of $f_i$'s moments, from the continuum velocity space to the discrete one. To do so, a Gauss-Hermite quadrature is applied~\cite{SHAN_JFM_550_2006,NIE_EPL_81_2008}:
\begin{equation}
	\int \boldsymbol{\mathcal{H}}^{(M)} f^{(0),N} \mathrm{d}\boldsymbol{\xi} = \sum_{i=1}^{V} \boldsymbol{\mathcal{H}}_i^{(M)} f_i^{(0),N}\ \mathrm{if}\ M+N \leq Q.
\label{eq:HermiteOrthoPreservation}	
\end{equation}
Here $\boldsymbol{\mathcal{H}}_i^{(M)}=\boldsymbol{\mathcal{H}}^{(M)}(\boldsymbol{\xi}_i)$, $f_i^{(0),N}=\frac{\omega_i}{\omega(\boldsymbol{\xi}_i)}f^{(0),N}(\boldsymbol{\xi}_i)$, $\omega_i$'s are the gaussian weights of the quadrature, and $Q$ is the order of accuracy of the quadrature. The resolution of this quadrature problem aims at finding lattices, \emph{i.e.}, discrete weights $\omega_i$'s associated with discrete velocities $\boldsymbol{\xi_i}$'s, allowing the conservation of Hermite polynomials orthogonality properties up to a requested order $M$. Then, the $M$th-order moment can be exactly recovered if the equilibrium function is truncated to the order $N\geq M$, and if the quadrature order is greater than $2N-1$. Some common two- and three-dimensional lattices, satisfying Eq.~(\ref{eq:HermiteOrthoPreservation}) for several $N$, are detailed in App.~\ref{sec:appendixD}. 

In the rest of the paper, the regularization procedure will be assessed on three lattices: the common D2Q9 lattice and the high-order D2V17 and D2V37 lattices. The D2V37 lattice, for which $Q=9$, allows the conservation of moments up to $M=4$. Therefore, this lattice is able to reproduce the macroscopic behavior of the fully compressible set of Navier-Stokes-Fourier equations. On the contrary, the D2V17 lattice flows from a seventh-order quadrature, which means that the equilibrium VDF can be truncated up to the third order only, leading to truncation error terms in the energy equation~(\ref{eq:errorN3}). This lattice should then be restricted to isothermal flows, even though the simulation of flows with weak temperature fluctuations could also be valid. Eventually, the case of the D2Q9 lattice, with a quadrature order of $Q=5$, allows developments up to the second order only, which leads to the famous compressibility error in the momentum equation~(\ref{eq:errorN2}).  

\subsection{\label{sec:PRvsRR}Regularization step: Projection vs Recursivity}
 
In the general case of a lattice including $V$ discrete speeds, the dimension of the associated Hilbert space is also $V$. Assuming the quadrature order of this lattice structure is $Q$, then Hermite polynomials up to the order $N=(Q-1)/2$ will be orthogonal to each other~\cite{NIE_EPL_81_2008} and may form part of a basis $\mathcal{B}= \mathcal{B^{\mathcal{H}}} \cup \mathcal{B^{\overline{\mathcal{H}}}}$, where
\begin{equation}
\mathcal{B^{\mathcal{H}}} = \left(\bm{\mathcal{H}}_i^{(0)},\ldots,\bm{\mathcal{H}}_i^{(N)}\right)
\end{equation}
is a sub-set of $\mathcal{B}$ entirely composed of Hermite polynomials, while elements of $\mathcal{B^{\overline{\mathcal{H}}}}$ are linearly independent of each others and may not be Hermite polynomials. Using this decomposition, the polynomial coefficients of $f^{(1)}$ can also be recast into two sub-sets:
\begin{equation}
	\left\{\boldsymbol{a}_1^{(0)},\ldots,\boldsymbol{a}_1^{(N)}\right\}^{\mathcal{B}^{\mathcal{H}}} \:\:\&\quad\:\: \left\{\bm{b}_1^{(n)}\right\}^{\mathcal{B}^{\overline{\mathcal{H}}}}_{n > N},
\end{equation}
where coefficients $\bm{b}_1^{(n)}$ ($n > N$) are supposedly related to non-hydrodynamic behaviors. 

The purpose of the PR approach is to keep only Hermite polynomial coefficients, since the mathematical expression of $\bm{b}_1^{(n)}$ ($n > N$) is usually unknown. After the PR procedure, remaining polynomial coefficients of $f^{(1),N}_{i,PR}$ are
\begin{equation}
\left\{\boldsymbol{a}_{1,PR}^{(0)},\ldots,\boldsymbol{a}_{1,PR}^{(N)}\right\}^{\mathcal{B}^{\mathcal{H}}},
\end{equation}
where contributions from $\mathcal{B}^{\overline{\mathcal{H}}}$ have been completely filtered out. Nevertheless, spurious sources coming from the approximation $f_i^{(1)}\approx (f_i - f_i^{(0)})$ may still be hidden in $\boldsymbol{a}_{1,PR}^{(n)}$.
Hence, this approach reduces the order of the polynomial development, and filters out spurious contributions originating from presumably non-Hermite polynomials. In the particular case of $\mathcal{B}^{\overline{\mathcal{H}}}$ being empty, this regularization step reduces to the standard BGK collision model if $f_i^{(1)}$ is projected onto the \emph{complete} basis $\mathcal{B}$. We will see later that $\mathcal{B}_{D2Q9}$ and $\mathcal{B}_{D3Q27}$ belong to this particular case.

Regarding now the RR approach, it further filters out high-order contributions, left by the PR approach, recomputing most coefficients $\bm{a}_1^{(n)}$ ($n\leq N$) by a Chapman-Enskog expansion, and without assuming $f^{(1)}_i\approx (f_i - f^{(0)}_i)$. For $f^{(1),N}_{i,RR}$ the remaining coefficients are then
\begin{equation}
\left\{\boldsymbol{a}_{1,RR}^{(0)},\ldots,\boldsymbol{a}_{1,RR}^{(N)}\right\}^{\mathcal{B}^{\mathcal{H}}},
\end{equation}
where $\boldsymbol{a}_{1,RR}^{(n)} = \boldsymbol{a}_{1,PR}^{(n)}$ for $n \leq 2$ ($n \leq 3$) in the isothermal (thermal) case, whereas high-order coefficients are recomputed using Eqs.~(\ref{eq:MalaspinasRecursiveRelation}) and~(\ref{eq:RecursiveRelation_ThermalCase}).
Eventually, while working on a \emph{complete} basis, this approach is the only one leading to the expected filtering behavior.

For the velocity sets of interest, the following observations can be made. The D2V37 (D2V17) is built ensuring that Hermite polynomials orthogonality properties are preserved, up to $N=4$ ($N=3$), during the velocity space discretization~\cite{PHILIPPI_PRE_73_2006}. Thus it is known for sure that
 \begin{align}
  \mathcal{B}_{D2V37}^{\mathcal{H}} = \Big(&\mathcal{H}^{(0)},\mathcal{H}^{(1)}_x,\mathcal{H}^{(1)}_y, \mathcal{H}^{(2)}_{xx},\mathcal{H}^{(2)}_{yy},\mathcal{H}^{(2)}_{xy},\nonumber\\[0.1cm]
  &\mathcal{H}^{(3)}_{xxx},\mathcal{H}^{(3)}_{yyy},\mathcal{H}^{(3)}_{xxy},\mathcal{H}^{(3)}_{xyy},\mathcal{H}^{(4)}_{xxxx},\\[0.1cm]
  &\mathcal{H}^{(4)}_{yyyy},\mathcal{H}^{(4)}_{xxxy},\mathcal{H}^{(4)}_{xyyy},\mathcal{H}^{(4)}_{xxyy}\Big), \nonumber 
  \end{align}
and  
 \begin{align}
  \mathcal{B}_{D2V17}^{\mathcal{H}} = \Big(&\mathcal{H}^{(0)},\mathcal{H}^{(1)}_x,\mathcal{H}^{(1)}_y,\mathcal{H}^{(2)}_{xx},\mathcal{H}^{(2)}_{yy},\mathcal{H}^{(2)}_{xy},\nonumber\\[0.1cm]
  &\mathcal{H}^{(3)}_{xxx},\mathcal{H}^{(3)}_{yyy},\mathcal{H}^{(3)}_{xxy},\mathcal{H}^{(3)}_{xyy}\Big),  
  \end{align}
whereas the true form of $\mathcal{B}_{D2V37}^{\overline{\mathcal{H}}}$ and $\mathcal{B}_{D2V17}^{\overline{\mathcal{H}}}$ are unknown. Hence, both PR and RR approaches first discard coefficients related to $\mathcal{B}_{D2V37}^{\overline{\mathcal{H}}}$ and $\mathcal{B}_{D2V17}^{\overline{\mathcal{H}}}$. Then, $\boldsymbol{a}_{1}^{(n)}$ ($n\leq N$) are computed either by the PR or by the RR approach.
In the particular case of the D2Q9 lattice, Hermite polynomials up to $N=2$ were first considered for the polynomial expansion~\cite{QIAN_EPL_17_1992}, leading to
\begin{align}
	\mathcal{B}^{\mathcal{H}}_{D2Q9}=\Big( &\mathcal{H}_i^{(0)}, \mathcal{H}_{i,x}^{(1)}, \mathcal{H}_{i,y}^{(1)}, \mathcal{H}_{i,xx}^{(2)}, \mathcal{H}_{i,xy}^{(2)}, \mathcal{H}_{i,yy}^{(2)}\Big).
\end{align}
Nevertheless, the development of the VDF can be extended including some third- and fourth-order terms, which also satisfy the orthogonality property conservation~(\ref{eq:HermiteOrthoPreservation}):
\begin{equation}
\mathcal{B}^{\mathcal{H},Complete}_{D2Q9}=\mathcal{B}^{\mathcal{H}}_{D2Q9} \cup \Big(\mathcal{H}_{i,xxy}^{(3)}, \mathcal{H}_{i,xyy}^{(3)}, \mathcal{H}_{i,xxyy}^{(4)}  \Big)
\end{equation}
and $\mathcal{B}^{\overline{\mathcal{H}},Complete}_{D2Q9}= \O$. In other words, the \emph{complete} basis $\mathcal{B}_{D2Q9}$ can be derived thanks to the tensor properties of the D2Q9 lattice, which allows the orthogonality conservation of every second-order Hermite polynomial \emph{per direction}. The same applies to the D3Q27 lattice.\\ 
Malaspinas has recently shown that developing $f_i^{(0),N}$ onto these \emph{complete} basis helps reducing the $\mathcal{O}(Ma^3)$ error term in the Navier Stokes' equations by removing all non-diagonal terms~\cite{MALASPINAS_ARXIV_2015}. The \emph{same} development was also done for $f_i^{(1),N}$ during the regularization process. And even if it was not explicitly specified in~\cite{MALASPINAS_ARXIV_2015}, one should notice that doing a full projection of the non-equilibrium VDF onto $\mathcal{B}_{D2Q9}$, in order to compute the 9 off-equilibrium coefficients, has no impact at all on the off-equilibrium VDF. As an analogy, it would be pointless to project a 2D vector onto the orthonormal basis of the 2D physical space using the Euclidean scalar product. Hence a \emph{complete} PR approach would be absolutely useless, since it would not filter out any physical information.

The rest of the paper will aim at showing some interesting properties of the RR procedure:
\begin{enumerate}
	\item It enhances numerical stability compared to the PR procedure for both standard and high-order LBMs,
	\item It is the only way to filter out non-hydrodynamic spurious sources without discarding any Hermite coefficients,
	\item The extension of the RR collision model~(\ref{eq:RecursiveRelation_ThermalCase}) helps improving numerical stability for the simulation of thermal and fully compressible flows.
\end{enumerate}

\section{\label{sec:NumericalAssessment}NUMERICS \& VALIDATION }
In this section, the principle of the space/time discretization of the LBE is first recalled. Then, improvements regarding numerical stability and  accuracy, induced by the use of the RR collision model, are confirmed for both standard and high-order LBMs. For the sake of clarity, the truncation notation $\cdot^N$ is dropped.
\subsection{\label{sec:NumericalDiscretization}Key points on the space/time discretization}
Let us start from the force-free lattice-Boltzmann equation with a general collision operator,
\begin{equation}
\partial_t f_i + \boldsymbol\xi_i\cdot\boldsymbol\nabla f_i = \Omega_i.
\label{eq:GeneralLBE} 
\end{equation} 
This equation is a first-order partial differential equation. The LHS term is \emph{linear} and corresponds to the convection of the discrete VDFs $f_i$'s at \emph{constant} speed $\xi_i$, while the RHS term is \emph{non-linear} and translates the rate of change of $f_i$'s induced by collisions. Since these two contributions behave differently from the mathematical point of view, two different time integrations are used: (a) the method of characteristics allows to \emph{exactly} integrate the convection term between $t$ and $t+\Delta t$, whereas (b) the trapezoidal rule ensures a \emph{second-order} accuracy in time for the collision term integration. These techniques lead to
\begin{align}
f_i\left(\boldsymbol{x} + \boldsymbol\xi_i\Delta t, t + \Delta t\right) - f_i\left(\boldsymbol{x}, t\right) &= \dfrac{ \Omega_i\left(t + \Delta t\right)+ \Omega_i(t)}{2}\Delta t \nonumber\\[0.1cm]
&+ \mathcal{O}(\Delta t^2, \Delta x^2),
\label{eq:implicitTimeDis}
\end{align}
where the space discretization error $\mathcal{O}(\Delta x^2)$ comes from (a): $\Delta \boldsymbol{x} = \boldsymbol{\xi_i}\Delta t$. This space/time discretization results in an implicit formulation since $\Omega_i\left(t + \Delta t\right)$ depends on $f\left(\boldsymbol{x} + \boldsymbol\xi_i\Delta t, t + \Delta t\right)$. Nevertheless, a change of variables compliant with the conservation of mass, momentum and total energy allows to get around it~\cite{DELLAR_PRE_64_2001}
\begin{equation}
\overline{f_i}\left(\boldsymbol{x}, t\right) =  f_i\left(\boldsymbol{x}, t\right) - \dfrac{\Delta t}{2}\Omega_i(t),
\label{eq:changeOfVariables}
\end{equation}
and leads to an explicit numerical scheme,
\begin{equation}
\overline{f_i}\left(\boldsymbol{x} + \boldsymbol\xi_i\Delta t, t + \Delta t\right) - \overline{f_i}\left(\boldsymbol{x}, t\right) = \Delta t \,\Omega_i(t) .
\end{equation}
To be fully consistent with the new set of VDFs $\overline{f_i}$'s, the collision operator needs to be slightly modified. For the BGK collision model,
\begin{align}
\Delta t\,\Omega_i(t) &= -\dfrac{\Delta t}{\tau}\left[ f_i\left(\boldsymbol{x}, t\right) -  f^{(0)}_i\left(\boldsymbol{x}, t\right)\right] \nonumber\\[0.1cm]
                    &= -\dfrac{\Delta t}{\tau}\left[ \overline{f_i}\left(\boldsymbol{x}, t\right) + \dfrac{\Delta t}{2}\Omega_i(t) -  f^{(0)}_i\left(\boldsymbol{x}, t\right)\right] \nonumber\\[0.1cm]
                    &= -\dfrac{1}{\overline{\tau}}\left[ \overline{f_i}\left(\boldsymbol{x}, t\right) -  f^{(0)}_i\left(\boldsymbol{x}, t\right)\right].
\end{align}
with $\overline{\tau} = \tau/\Delta t \,+\, 1 /2$. Dropping the overline notation, and adopting the shorthand notation $\Delta x = \Delta t = 1$, we end up with the famous second-order accurate and explicit numerical scheme
\begin{equation}
f_i\left(\boldsymbol{x} +1, t + 1\right) - f_i\left(\boldsymbol{x}, t\right) = -\dfrac{\left[ f_i\left(\boldsymbol{x}, t\right) -  f^{(0)}_i\left(\boldsymbol{x}, t\right)\right]}{\tau}.
\end{equation}
All results presented hereafter will be based on the coupling between this numerical scheme and several discretizations of the velocity space, namely, the D2Q9, the D2V17 and the D2V37 velocity sets. These lattice structures were built using an on-grid condition~\cite{PHILIPPI_PRE_73_2006},
\begin{equation}
\boldsymbol\xi_i \longrightarrow \dfrac{1}{c_s}\boldsymbol{\xi}_i,\quad\vert\vert\boldsymbol{\xi}_i\vert\vert\in\mathbb{N},
\end{equation}
allowing to achieve the full potential of the method of characteristics. The renormalization constant (or lattice constant) $c_s$ can further be identified as the isothermal lattice speed of sound, \emph{i.e.}, $c_s = \sqrt{rT_0}\Delta t/\Delta x$~\cite{VIGGEN_PRE_90_2014}. Its value is recalled for each lattice structure in App.~\ref{sec:appendixD}. In the rest of the paper, all necessary quantities used to define each numerical case will be expressed in lattice and dimensionless units.

\subsection{\label{sec:IsothermalValidation}Isothermal LBM ($\theta = 1$)}
In a first time, the case of isothermal flows is considered. Lattice structures with increasing complexity (D2Q9, D2V17 and D2V37) are successively used to point out the capability of the RR procedure to deal with both standard and high-order LBMs in a straightforward way. All velocity set definitions are recalled in App.~\ref{sec:appendixD}. 

\begin{figure}[ht!]
\begin{tabular}{c}
\includegraphics[width=0.44\textwidth,trim=0cm 2.cm 0cm 0cm,clip='true']{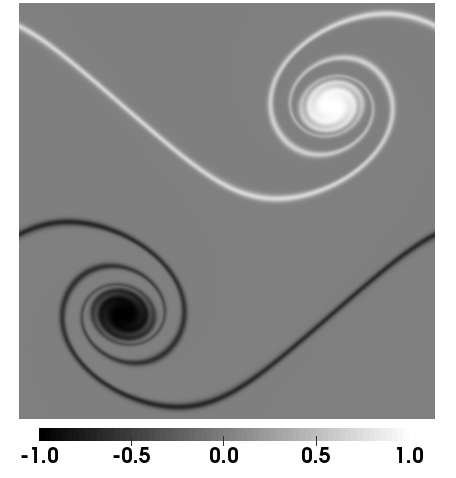}\\
\includegraphics[width=0.42\textwidth]{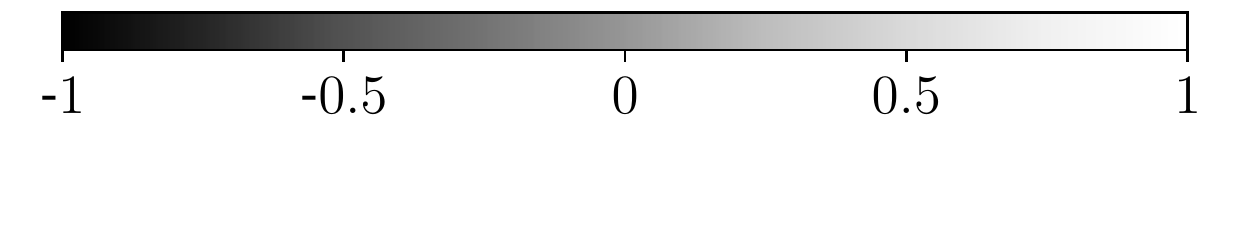}\\[-0.75cm]
\end{tabular}
\caption{Roll-up of the double shear layer at $M_0=0.2$ and $Re = 3\times 10^4$. Visualization of the dimensionless vorticity field at $t_c = L/ u_0 = 1$, using $L=512$ grid points in each directions, with the D2Q9 and the BGK collision model.}
\label{fig:KH_Instability}
\end{figure}

The double shear layer is a well-known test case which allows to quantify the stability of numerical schemes as a first step~\cite{MINION_JCP_138_1997}. This flow is composed of two longitudinal shear layers, located at $y=L/4$ and $y=3L/4$, in a 2D doubly periodic domain with $(x,y)\in [0,L]^2$. Furthermore, a transverse perturbation is superimposed to the flow. This leads to the roll-up of the shear layers, and the generation of two counter-rotating vortices by the Kelvin-Helmholtz instability mechanism. In addition, any numerically induced disturbances may lead to the formation of further spurious vortices, or in the worst case, make the simulation reach its stability threshold. This is why this test case is an excellent candidate to evaluate the stability of numerical schemes. The initial state is defined by,
\begin{equation}
u_x = \left\{
\begin{array}{l}
u_0\tanh[k(y^*-1/4)], \quad y^*\leq 1/2 \\[0.1cm]
u_0\tanh[k(3/4-y^*)], \quad y^*>1/2
\end{array} 
\right.
\label{eq:MB_InitialState_ux}
\end{equation}
and
\begin{equation}
u_y = u_0\delta \sin[2\pi(x^*+1/4)],
\label{eq:MB_InitialState_uy}
\end{equation}
where $(x^*,y^*) = (x/L,y/L)$, and $u_0$ is the characteristic speed. $k$ is related to the width of the shear layers while $\delta$ controls the amplitude of the transverse perturbation. Here the case of thin shear layers, where $(k,\delta)=(80,0.05)$, is considered. The Reynolds number, which is the ratio between convective and diffusive phenomena, is first fixed to a moderate value of $Re = u_0 L /\nu = 3\times 10^4$ (cf Fig.~\ref{fig:KH_Instability}). This is sufficient to reach the stability limit of the BGK collision model for standard LBMs when an under-resolved mesh ($L = 128$) is considered~\cite{BOSCH_ESAIM_52_2015}. 
The stability range of the proposed model is then evaluated varying the Reynolds number from $10^4$ to $10^6$. Simulations using the D2Q9, the D2V17 and the D2V37 lattices are performed with a free-stream Mach number $M_0 = u_0/c_s = 0.2,$ $0.35$ and $0.57$ respectively. These values are chosen in order to: (\emph{i}) properly distinguish the impact of each collision operator on the numerical stability of each LBM, and (\emph{ii}) reduce the impact of the $\mathcal{O}(Ma^3)$ error encountered when the D2Q9 lattice is employed. The initialization step is achieved using the approximation $f_i \approx f_i^{(0)} + f_i^{(1)}$, with $f_i^{(1)}$ computed using analytic formulas for the velocity gradients. This allows to reduce spurious oscillations at the beginning of the simulation as demonstrated in~\cite{MATTILA_PRE_91_2015}. Finally, it should be noted that for all simulations below, $f_i^{(0)}$ is expanded to the maximal authorized order, \emph{i.e.}, $N=3$ for the D2V17, and $N=4$ for both the D2Q9 and the D2V37, using their associated Hermite tensor basis (see App.~\ref{sec:appendixC} \&~\ref{sec:appendixD} for more details on this last point). 

\begin{figure}[t!]
\includegraphics[width=0.45\textwidth]{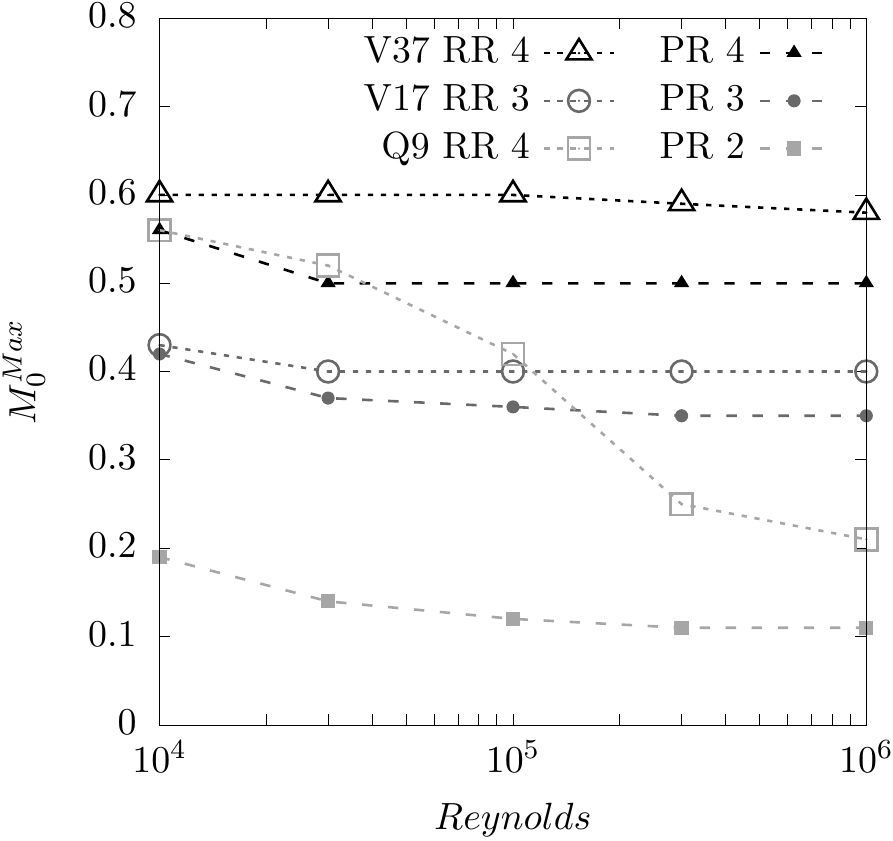}
\caption{Stability range of the double shear layer simulation using $L=128$, for $t \leq 2t_c$, and with the characteristic time $t_c = L / M_0 c_s$. The standard PR (filled symbols) and the most stable RR (open symbols) procedures are compared for every LBM. The RR approach always shows a higher stability range regarding $M_0^{Max}$, where the level of accuracy of $M_0^{Max}$ is $\Delta M_0 = 0.01$.}
\label{fig:DSL_StabilityRange_N128}
\end{figure}

Extensive results concerning the case at $Re=3\cdot10^4$ are compiled in Figs.~\ref{fig:DSL_Re3e04_M02_D2Q9_fEqO4},~\ref{fig:DSL_Re3e04_M035_D2V17_fEqO3} and~\ref{fig:DSL_Re3e04_M057_D2V37_fEqO4} for the D2Q9, the D2V17 and the D2V37 lattices respectively. All models are compared to a reference solution obtained using the BGK collision model with $L=2048$. Several partial conclusions can be drawn from these results.\\
\noindent Firstly, correct kinetic-energy-related evolutions are recovered for all computations, even using a relatively coarse mesh ($L=128$). Regarding mean and standard deviation of the enstrophy, a convergence study has been conducted, leading to the choice of a centered and fourth-order-accurate finite-difference scheme for the gradient evaluation. Nevertheless small errors are still observed, when the Mach number is increased, even for $L=256$. Thus recovering the proper evolution of the mean and the standard deviation of the enstrophy is more difficult than obtaining the correct evolution of the same quantities in the case of the kinetic energy. The study of the enstrophy evolution should then be preferred for the accuracy evaluation of LBMs.\\
\noindent Secondly, the new RR approach is more stable than the PR one, at least for the present lattices. As explained in Sec.~\ref{sec:PRvsRR}, this originates from a better computation of non-equilibrium Hermite coefficients $\boldsymbol{a}_1^{(n)}$ from the kinetic theory point of view. This point is further highlighted in Fig.~\ref{fig:DSL_StabilityRange_N128}, where the maximal achievable Mach number $M_0^{Max}$ allowing a stable simulation, up to $t/t_c = 2$, is plotted for Reynolds numbers ranging from $10^4$ to $10^6$, in the under-resolved configuration $L=128$.

\begin{widetext}

\begin{figure}[h!]
\centering
\includegraphics[width=0.9\textwidth]{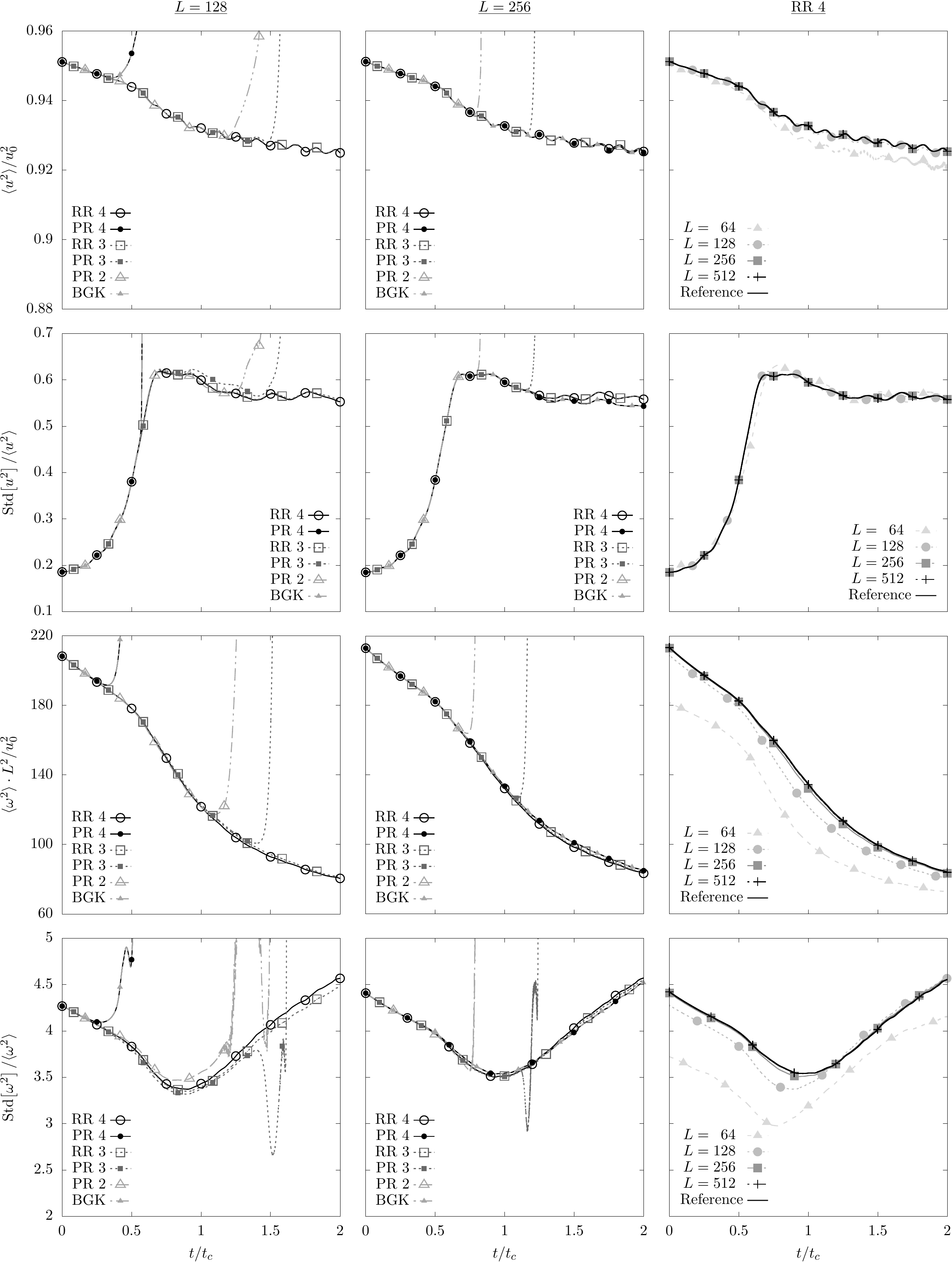}
\caption{Double shear layer at $M_0=0.2$ and $Re = 3\times 10^4$ for the D2Q9. From top to bottom: dimensionless mean kinetic energy, standard deviation of the kinetic energy, mean enstrophy and standard deviation of the enstrophy. All quantities are spatially averaged over all the simulation domain. The projection-based (PR) and the recursive (RR) regularization, at order $N = 2$, $3$ and $4$, are compared against the standard collision model (BGK). The first two columns are the results obtained using a $L\times L$ mesh with $L=128$ and $L=256$ respectively. The last column illustrates the mesh convergence of the fourth-order recursive regularization (RR 4), where the reference solution was obtained using the BGK collision operator with $L=2048$. The characteristic time $t_c$ is defined as $t_c = L/ u_0$, while the characteristic speed is $u_0=M_0 c_s$.}
\label{fig:DSL_Re3e04_M02_D2Q9_fEqO4}
\end{figure}\clearpage

\begin{figure}[h!]
\centering
\includegraphics[width=0.9\textwidth]{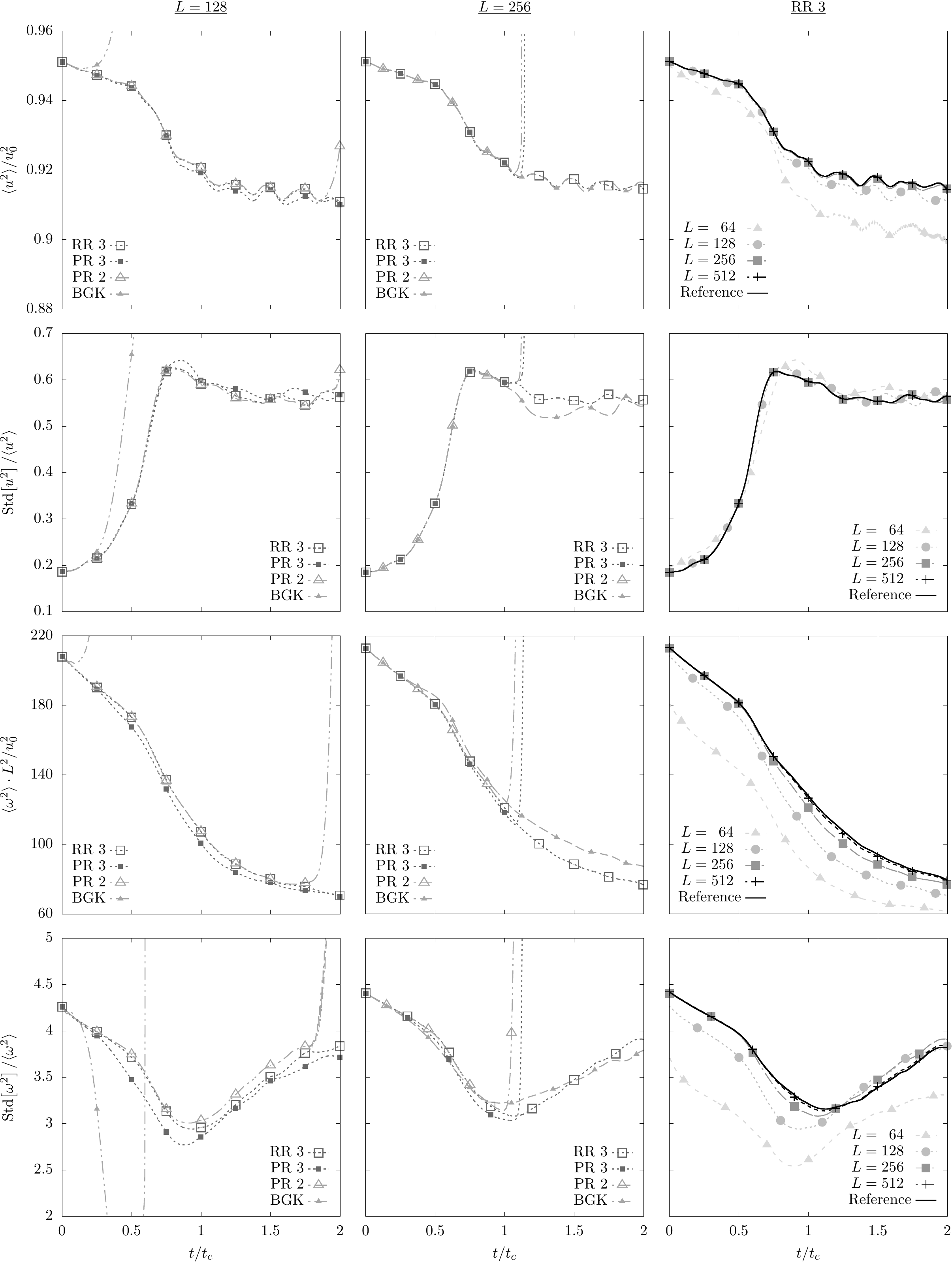}
\caption{Double shear layer at $M_0=0.35$ and $Re = 3\times 10^4$ for the D2V17. From top to bottom: dimensionless mean kinetic energy, standard deviation of the kinetic energy, mean enstrophy and standard deviation of the enstrophy. All quantities are spatially averaged over all the simulation domain. The projection-based (PR) and the recursive (RR) regularization, at order $N = 2$ and $3$, are compared against the standard collision model (BGK). The first two columns are the results obtained using a $L\times L$ mesh with $L=128$ and $L=256$ respectively. The last column illustrates the mesh convergence of the third-order recursive regularization (RR 3), where the reference solution was obtained using the BGK collision operator with $L=2048$. The characteristic time $t_c$ is defined as $t_c = L/ u_0$, while the characteristic speed is $u_0=M_0 c_s$.}
\label{fig:DSL_Re3e04_M035_D2V17_fEqO3}
\end{figure}\clearpage

\begin{figure}[h!]
\centering
\includegraphics[width=0.9\textwidth]{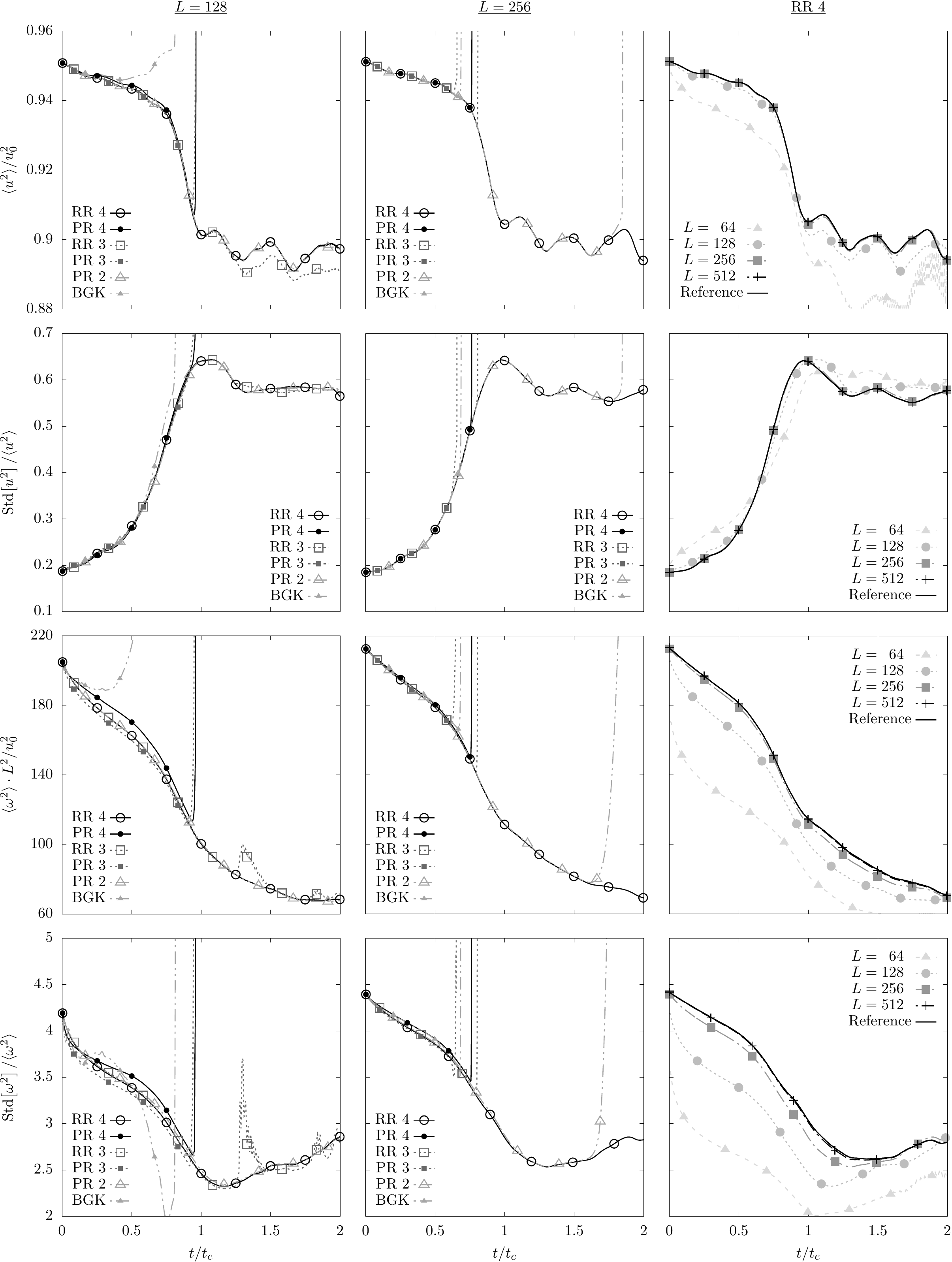}
\caption{Double shear layer at $M_0=0.57$ and $Re = 3\times 10^4$ for the D2V37. From top to bottom: dimensionless mean kinetic energy, standard deviation of the kinetic energy, mean enstrophy and standard deviation of the enstrophy. All quantities are spatially averaged over all the simulation domain. The projection-based (PR) and the recursive (RR) regularization, at order $N = 2$, $3$ and $4$, are compared against the standard collision model (BGK). The first two columns are the results obtained using a $L\times L$ mesh with $L=128$ and $L=256$ respectively. The last column illustrates the mesh convergence of the fourth-order recursive regularization (RR 4), where the reference solution was obtained using the BGK collision operator with $L=2048$. The characteristic time $t_c$ is defined as $t_c = L/ u_0$, while the characteristic speed is $u_0=M_0 c_s$.}
\label{fig:DSL_Re3e04_M057_D2V37_fEqO4}
\end{figure}\clearpage

\end{widetext}

\noindent The stability criterium that has been chosen is based on the mean kinetic energy $\langle u^2\rangle$: a computation is considered to remain stable if $\langle u^2(t\leq 2 t_c)\rangle < \langle u^2(t=0)\rangle$. For a proper comparison, the standard PR and the most stable RR versions associated to each lattice structure are compared. Results obtained in the particular case of $Re = 3\times 10^4$ seem to be extendable to a wide range of Reynolds number.\\
\noindent Thirdly, the PR 4 and the BGK collision models give exactly the same results in the particular case of the D2Q9. This confirms what was anticipated in Sec.~\ref{sec:PRvsRR}, \emph{i.e.}, when the VDF is expanded over the \emph{complete} Hermite basis then the PR 4 and the BGK reduce to the same collision operator. Therefore the only way to reach the full potential of the D2Q9 lattice is to use the RR procedure.

\noindent Lastly, the computational overhead for our \emph{non-optimized} implementation is $1.5 \leq t_{RR}/t_{BGK} \leq 2$, the lower limit is for the D2Q9 and the upper for the D2V37. This is far from being excessive considering the tremendous improvements obtained in terms of numerical stability. Regarding the regularization step itself, the RR approach is always faster (about $20 \%$) than the standard PR one. 

\subsection{\label{sec:FCompressibleValidation}Fully compressible LBM} 

To further validate the general RR process, the simulation of thermal and fully compressible flows is now considered. This is done through the numerical computation of the famous Sod shock tube~\cite{SOD_JCP_27_1978}, using the most compact lattice structure allowing to ensure the preservation of the orthogonality of all fourth-order Hermite tensors: the D2V37 velocity set~\cite{PHILIPPI_PRE_73_2006,SHAN_JCS_17_2016}. For this purpose, Hermite coefficients $\boldsymbol{a}_0^{(n)}$ and $\boldsymbol{a}_1^{(n)}$ are computed thanks to our proposed extensions~(\ref{eq:RecursivityEquilibrium}) and~(\ref{eq:RecursiveRelation_ThermalCase}).\\
This 1D Riemann problem consists in a closed tube divided into two regions by a thin membrane. Each region is filled with the same gas but has different thermodynamic properties (density $\rho$, temperature $T$, pressure $P$ and velocity $u$). At the initialization, the breakdown of the membrane induces a strong acceleration of the flow, from the high-pressure side to the low-pressure one, whose purpose is to equalize the pressure inside the tube. This leads to the generation and the propagation of three characteristic waves: (1) the compression of the gas creates a \emph{shock} wave which propagates towards the low-pressure side, (2) the expansion of the gas towards the high-pressure side induces the propagation of the \emph{expansion} or \emph{rarefaction} wave, and (3) the separation between the two waves, namely, the \emph{contact discontinuity}. The latter can be seen as a fictitious diaphragm traveling at a constant speed towards the low-pressure side.

In this paper, two different configurations are studied. They share the same pressure ratio but differ when it comes to their temperature or density ones:
\begin{equation}
(P_L,\rho_L,u_L) = (10,8,0), \quad (P_R,\rho_R,u_R) = (1,1,0),
 \end{equation} 
 \begin{equation}
(P_L,\rho_L,u_L) = (10,2,0), \quad (P_R,\rho_R,u_R) = (1,1,0),
 \end{equation}  
where subscripts $L$ and $R$ stand for the left and the right states respectively. To avoid the contribution of the boundary conditions, the computation takes place in a periodic domain of length $2L_x$ centered around the location $x=L_x/2$ where the discontinuity between the two states belongs. The simulation domain is then spatially discretized using $L_x = 400$ grid points. Such a coarse mesh will allow to further highlight the numerical stability issues encountered computing discontinuities.

Results obtained using the PR and the RR processes, at order $N=4$, are plotted along $[0,L_x]$ in Fig.~{\ref{fig:SodShockTube_RegularizationImpact_BW}}. Even though both models are able to properly reproduce the generation and the propagation of all the characteristic waves of this 1D Riemann problem, the PR procedure introduces a coupling between high-order and Navier-Stokes physics in the form of standing waves, whereas the RR procedure completely filters them out even if small over/undershoots still remain. The latter can be attenuated using either a finer grid or a shock-capturing technique, such as a shock sensor~\cite{BOGEY_JCP_228_2009}. 

It must be noted that without regularization steps, the standard LBM encountered severe numerical stability issues for the present configurations, and could not be stabilized even using extremely fine meshes (more than $10000$ points in the longitudinal direction). This further points out that the RR collision operator allows to get more stable solutions without degrading the accuracy of the numerical scheme. 

\begin{widetext}

\begin{figure}
\centering
\begin{tabular}{c @{\hspace{1.cm}} c}
\includegraphics[width=0.45\linewidth]{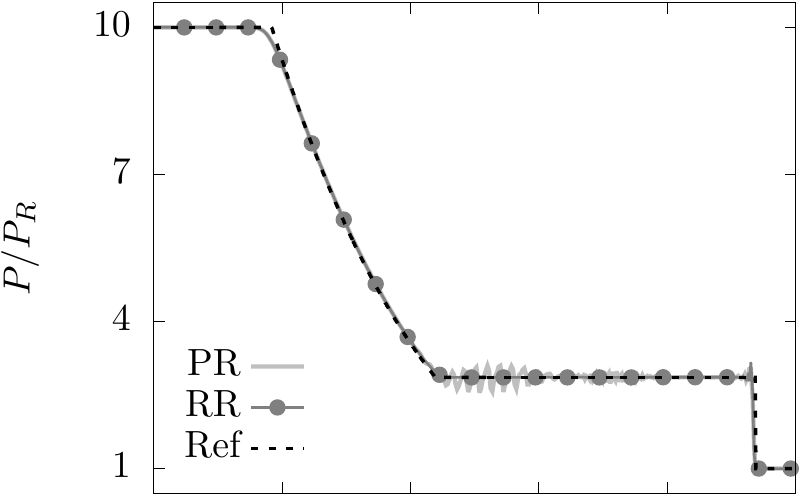} & \includegraphics[width=0.45\linewidth]{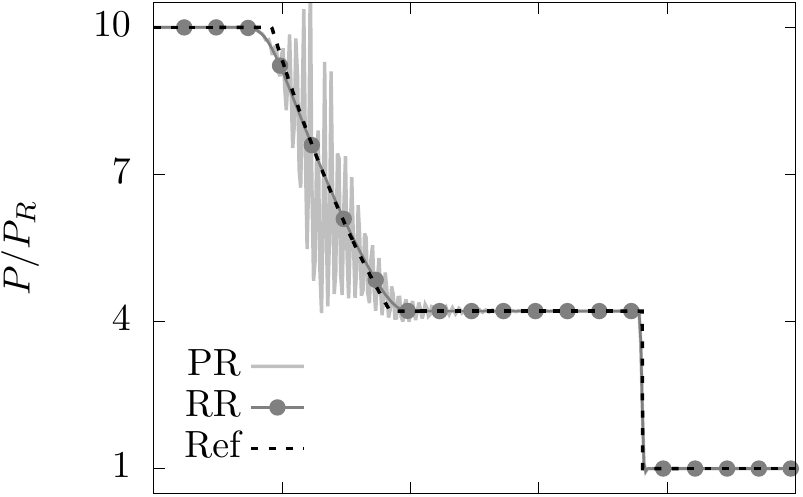}\\

\includegraphics[width=0.45\linewidth]{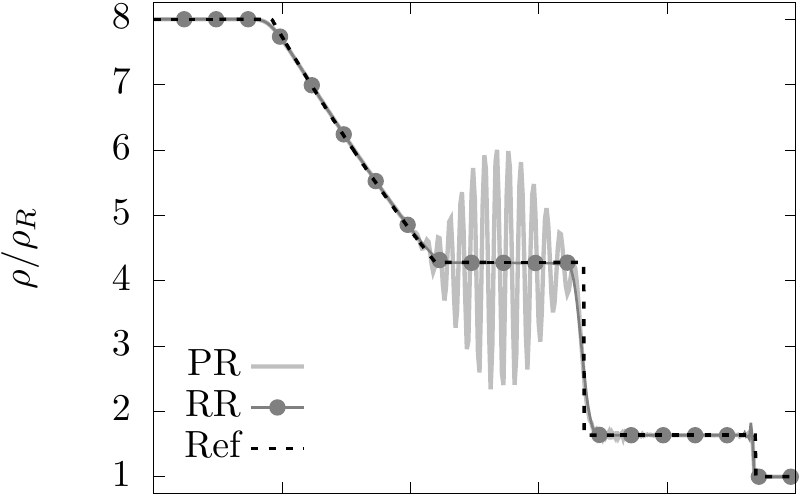} & \includegraphics[width=0.45\linewidth]{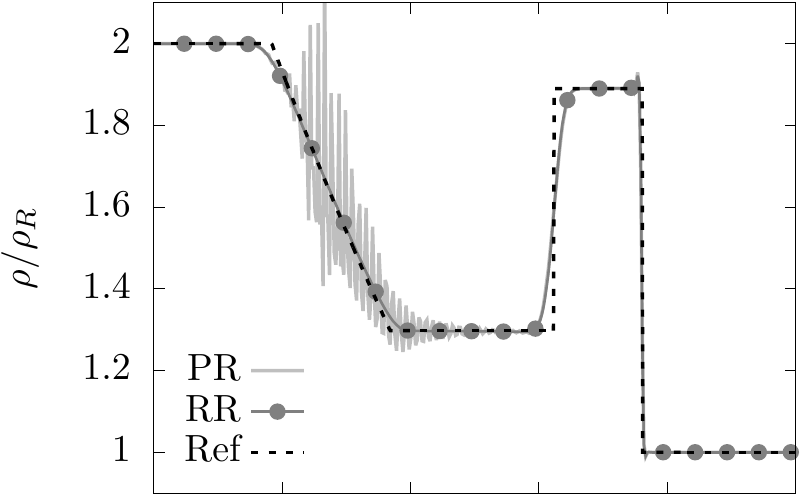}\\

\includegraphics[width=0.45\linewidth]{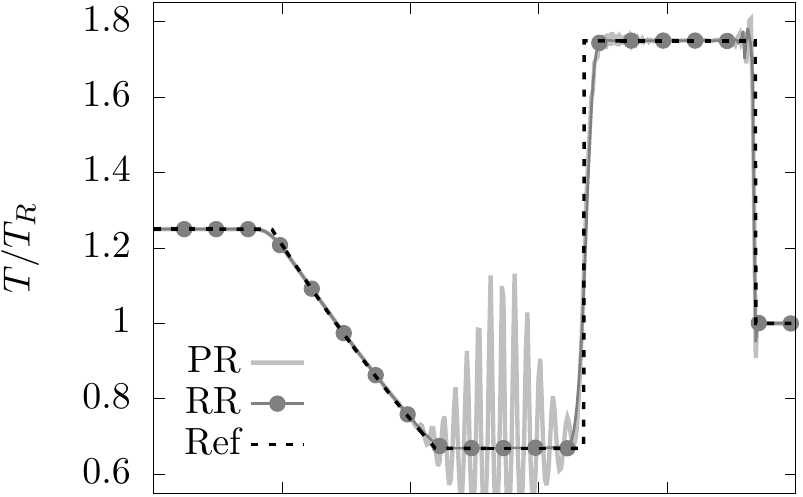} &
\includegraphics[width=0.45\linewidth]{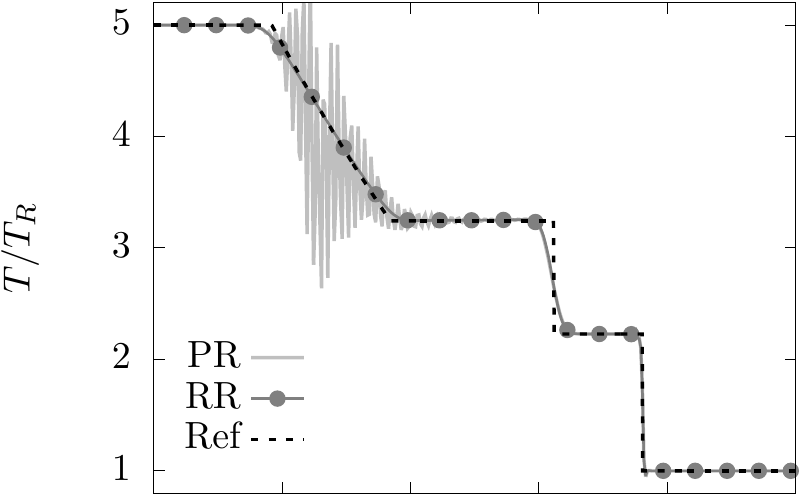}\\

\includegraphics[width=0.45\linewidth]{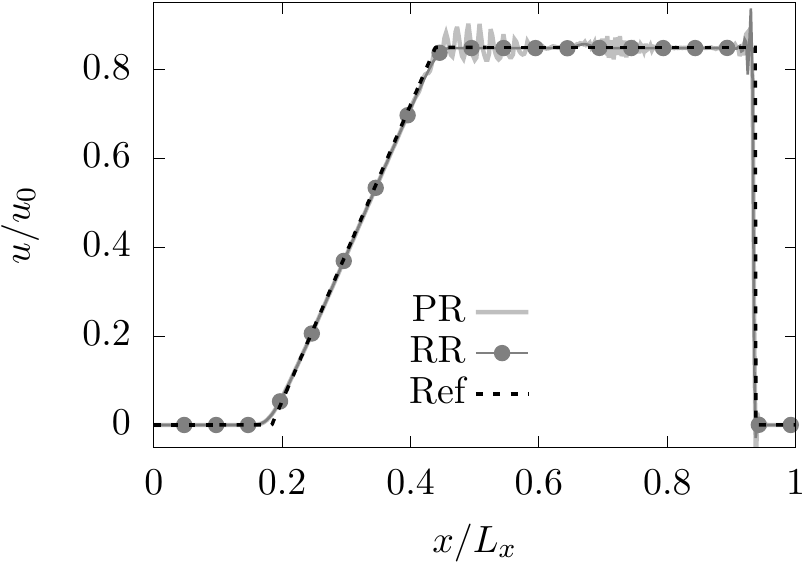} &
\includegraphics[width=0.45\linewidth]{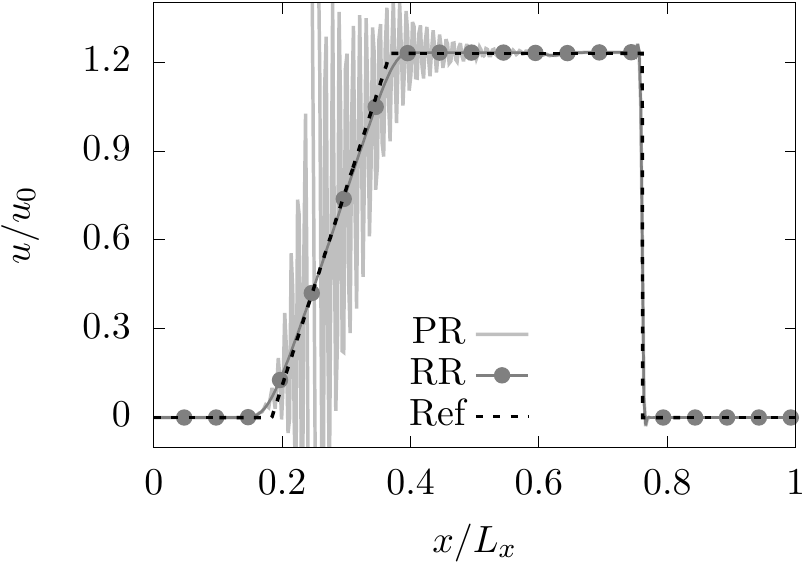} 
\end{tabular}
\caption{1D Riemann problem: $[P_L/P_R,\rho_L/\rho_R]=[10,8]$ (left) and $[10,2]$ (right) with $L_x=400$ grid points, and a relaxation time of $\tau = 0.595$ and $\tau = 0.760$ respectively. From top to bottom: dimensionless pressure, density, temperature and velocity profiles. Results obtained using the fourth-order PR and RR steps are compared to the reference solution (dashed line) for a specific heat ratio $\gamma = 2$. They are plotted at time $t/t_c = 0.2$ (left) and $t/t_c = 0.1$ (right), with the characteristic time $t_c=L_x/\sqrt{\gamma T_R}$.}
\label{fig:SodShockTube_RegularizationImpact_BW}
\end{figure}\clearpage

\end{widetext}

\section{\label{sec:Conclusion}CONCLUSION}
Despite a wide range of validity of the standard lattice Boltzmann method (LBM), the simulation of certain flows remains a tedious task: (1) weakly compressible flows at high Reynolds numbers, and (2) fully compressible flows including discontinuities such as shockwaves. In this context, the present work focuses on the extension of the recursive regularization step to high-order LBMs. New LBMs,  with increased stability range, are then obtained filtering out the non-hydrodynamic contributions induced by the streaming step. This technique relies on two points: (a) the computation of non-equilibrium coefficients $\boldsymbol{a}_1$ through the Chapman-Enskog expansion, eased thanks to (b) the recursive properties of Hermite polynomial basis. This procedure was originally derived in the particular case of isothermal and weakly compressible flows simulations, with standard lattice structures such as the D2Q9 and the D3Q27. Here we further validate it using high-order LBMs (the D2V17 and the D2V37) for the simulation of both isothermal flows and thermal fully compressible flows. The latter was possible thanks to our derivation of general recursive formulas for the computation of Hermite coefficients $\boldsymbol{a}_0$ and $\boldsymbol{a}_1$. Strong improvements in terms of numerical stability are confirmed for both kind of simulations, and at a relatively low computational overhead.

Flows including external accelerations, boundary conditions, and LBMs with two sets of populations are currently under investigation. Theoretical derivations and numerical validations, concerning the extension of the RR procedure to deal with these topics, will be presented in the near future.

\begin{acknowledgments}
C. C. wants to thank D. Ricot \& O. Malaspinas for fruitful conversations about the lattice Boltzmann theoretical background, and regularization steps. This work was supported by the French funded project CLIMB in the framework of the \emph{Programme d'Investissement et d'Avenir - Calcul Intensif et Simulation Num\'{e}rique}.
\end{acknowledgments}
  
\appendix

\onecolumngrid

\section{\label{sec:Recursivity_a1_proof}Proof of the recursive formula}

\noindent The aim of this section is to prove the following recursive relation:
\begin{align}
\label{eq:RecursiveRelation_ThermalCase_proof}
	\forall n \ge 4,\quad a_{1,{\alpha_1..\alpha_n}}^{(n)}&=u_{\alpha_{n}} a_{1,{\alpha_1..\alpha_{n-1}}}^{(n-1)} + c_s^2 (\theta-1) \sum_{i=1}^{n-1} \delta_{\alpha_i \alpha_n} a_{1,{\beta_i}}^{(n-2)} + \frac{1}{\rho} \sum_{i=1}^{n-1} a_{0,{\beta_i}}^{(n-2)} a_{1,{\alpha_i \alpha_n}}^{(2)} \nonumber\\
	&+ \frac{1}{\rho} \sum_{i=1}^{n-1} \sum_{j>i}^{n-1} a_{0,{\beta_{ij}}}^{(n-3)} \left( a_{1,{\alpha_i \alpha_j \alpha_n}}^{(3)} - u_{\alpha_i} a_{1,{\alpha_j \alpha_n}}^{(2)} - u_{\alpha_j} a_{1,{\alpha_i \alpha_n}}^{(2)} - u_{\alpha_n} a_{1,{\alpha_i \alpha_j}}^{(2)} \right).
\end{align}
where some mathematical notations are used for the sake of clarity: (1) $\beta_i$ is used when the index $\alpha_i$ is omitted, e.g., $a_{0,\beta_i}^{(n)} \equiv a_{0,\alpha_1..\alpha_{i-1}\alpha_{i+1}..\alpha_{n}}^{(n)}$, and (2) if $\alpha_i$ and $\alpha_j$ are omitted then $\beta_{ij}$ is used.\\\\
The following relations are needed to prove~(\ref{eq:RecursiveRelation_ThermalCase}):
\begin{itemize}
\item \emph{The Hermite coefficient based LBE}~\cite{SHAN_JFM_550_2006}
\begin{equation}
	\label{eq:LBK_ProjectedOnH}
	\forall n \ge 1,\quad  a_{1,{\alpha_1..\alpha_n}}^{(n)} = -\tau \left[ \partial_t a_{0,{\alpha_1..\alpha_n}}^{(n)} + \partial_{\gamma} a_{0,{\alpha_1..\alpha_{n}\gamma}}^{(n+1)} + c_s^2 \sum_{i=1}^{n} \partial_{\alpha_i} a_{0,{\beta_i}}^{(n-1)} \right],
\end{equation}
where Einstein summation notation is used on subscript $\gamma$,
\item \emph{The recursive formula for Hermite coefficients at equilibrium} (App.~\ref{sec:appendixB})
\begin{equation}
	\label{eq:Recursivity_a0}
	\forall n \ge 2,\quad a_{0,{\alpha_1..\alpha_n}}^{(n)} = u_{\alpha_n} a_{0,{\alpha_1..\alpha_{n-1}}}^{(n-1)} + (\theta-1) c_s^2 \sum_{i=1}^{n-1} \delta_{\alpha_i \alpha_n} a_{0,{\beta_i}}^{(n-2)},
\end{equation}
\item \emph{Euler's equations}
\begin{align}
&\partial_t \rho + \partial_\gamma (\rho u_\gamma) = 0,	\label{eq:Euler_mass}\\[0.2cm]
&\rho\partial_t (u_\alpha) +  \rho u_\gamma\partial_\gamma (u_\alpha) + \partial_\alpha p =0,	\label{eq:Euler_momentum}\\
&\partial_t \theta + u_\gamma \partial_\gamma \theta + \frac{2}{D}\theta \partial_\gamma u_\gamma = 0,\label{eq:Euler_temperature}
\end{align}
\end{itemize}
\noindent Let us now move on to the proof itself.\\\\
\underline{STEP 1: Boltzmann equation associated to Hermite coefficients (\ref{eq:LBK_ProjectedOnH}) at order \text{$(n-1)$}.}\\
\begin{align*}
	\forall n \ge 2,\quad  u_{\alpha_n} a_{1,{\alpha_1..\alpha_{n-1}}}^{(n-1)} &= -\tau \left[ u_{\alpha_n} \partial_t a_{0,{\alpha_1..\alpha_{n-1}}}^{(n-1)} + u_{\alpha_n} \partial_{\gamma} a_{0,{\alpha_1..\alpha_{n-1}\gamma}}^{(n)} + c_s^2 \sum_{i=1}^{n-1} u_{\alpha_n} \partial_{\alpha_i} a_{0,{\beta_i}}^{(n-2)} \right] \\
	&= -\tau \Bigg[ \partial_t \left( u_{\alpha_n} a_{0,{\alpha_1..\alpha_{n-1}}}^{(n-1)} \right) - a_{0,{\alpha_1..\alpha_{n-1}}}^{(n-1)} \partial_t u_{\alpha_n} + \partial_{\gamma} \left( u_{\alpha_n} a_{0,{\alpha_1..\alpha_{n-1}\gamma}}^{(n)} \right) \\
	&- a_{0,{\alpha_1..\alpha_{n-1}\gamma}}^{(n)} \partial_{\gamma} u_{\alpha_n} + c_s^2 \sum_{i=1}^{n-1} \partial_{\alpha_i} \left( u_{\alpha_n} a_{0,{\beta_i}}^{(n-2)} \right) - c_s^2 \sum_{i=1}^{n-1} a_{0,{\beta_i}}^{(n-2)} \partial_{\alpha_i} u_{\alpha_n} \Bigg].
\end{align*}
where the derivation by parts rule is used.
\\\\
\noindent\underline{STEP 2: Remove all time derivatives $\partial_t$.}\\\\
First, we use Eq.~(\ref{eq:Recursivity_a0}),
\begin{align*}
	u_{\alpha_n} a_{1,{\alpha_1..\alpha_{n-1}}}^{(n-1)} &=  -\tau \Bigg[ \underbrace{\partial_t a_{0,{\alpha_1..\alpha_n}}^{(n)} + \partial_{\gamma} a_{0,{\alpha_1..\alpha_n\gamma}}^{(n+1)} + c_s^2 \sum_{i=1}^n \partial_{\alpha_i} a_{0,{\beta_i}}^{(n-1)} }_{(i)} \\
	& \underbrace{- a_{0,{\alpha_1..\alpha_{n-1}}}^{(n-1)} \partial_t u_{\alpha_n} - a_{0,{\alpha_1..\alpha_{n-1}\gamma}}^{(n)} \partial_{\gamma} u_{\alpha_n} }_{(ii)} \\
	& \underbrace{ -c_s^2 \partial_{\alpha_n}a_{0,{\alpha_1..\alpha_{n-1}}}^{(n-1)} - c_s^2 \sum_{i=1}^{n-1} a_{0,{\beta_i}}^{(n-2)} \partial_{\alpha_i} u_{\alpha_n} }_{(iii)} \\
	& \underbrace{ -c_s^2 \sum_{i=1}^{n-1} \delta_{\alpha_i \alpha_n} \partial_t \left( (\theta-1)a_{0,{\beta_i}}^{(n-2)} \right) -c_s^2 \sum_{i=1,i\ne n}^{n+1} \delta_{\alpha_i \alpha_n} \partial_{\gamma} \left( (\theta-1)a_{0,{\beta_i\gamma}}^{(n-1)} \right) }_{(iv)} \\
	& \underbrace{-c_s^4 \sum_{i=1}^{n-1} \sum_{j=1,j\ne i}^{n-1} \delta_{\alpha_j \alpha_n} \partial_{\alpha_i} \left( (\theta-1) a_{0,{\beta_{ij}}}^{(n-3)} \right)  }_{(v)} \Bigg].
\end{align*}
for $n \ge 3$. Then, thanks to~(\ref{eq:LBK_ProjectedOnH}) one can replace $\partial_t a_{0,{\alpha_1..\alpha_n}}^{(n)}$ by
$$ (i)=-\frac{1}{\tau} a^{(n)}_{1,{\alpha_1..\alpha_n}} $$
and
$$ (ii) = -a_{0,{\alpha_1..\alpha_{n-1}}}^{(n-1)} (\partial_t u_{\alpha_n} + u_{\gamma} \partial_{\gamma} u_{\alpha_n}) - c_s^2 (\theta-1) \partial_{\gamma} u_{\alpha_n} \sum_{i=1}^{n-1} \delta_{\alpha_i \gamma} a_{0,{\beta_i}}^{(n-2)}. $$
In the same spirit as for Chapman-Enskog expansion, Euler's equation of momentum (Eq.~(\ref{eq:Euler_momentum})) is used to replace the time derivative $\partial_t u_{\alpha_n}$. In addition to that, removing the implicit summation on $\gamma$ leads to
$$ (ii)=a_{0,{\alpha_1..\alpha_{n-1}}}^{(n-1)} \frac{1}{\rho} \partial_{\alpha_n} p - c_s^2 (\theta-1) \sum_{i=1}^{n-1} a_{0,{\beta_i}}^{(n-2)} \partial_{\alpha_i} u_{\alpha_n}, $$
so that
$$ (ii) + (iii) = a_{0,{\alpha_1..\alpha_{n-1}}}^{(n-1)} \frac{1}{\rho} \partial_{\alpha_n}p -c_s^2\theta \sum_{i=1}^{n-1}a_{0,{\beta_i}}^{(n-2)} \partial_{\alpha_i} u_{\alpha_n} - c_s^2 \partial_{\alpha_n} a_{0,{\alpha_1..\alpha_{n-1}}}^{(n-1)}.$$
\noindent Let us now focus on $(iv)$ and $(v)$:
\begin{align*}
(iv) &= \underbrace{-c_s^2 \sum_{i=1}^{n-1} \delta_{\alpha_i \alpha_n} \left[ a_{0,{\beta_i}}^{(n-2)} \partial_t \theta + a_{0,{\beta_i\gamma}}^{(n-1)} \partial_{\gamma} \theta \right]}_{(iv.i)}  \underbrace{-c_s^2 (\theta-1) \sum_{i=1}^{n-1} \delta_{\alpha_i \alpha_n} \left[ \partial_t a_{0,{\beta_i}}^{(n-2)} + \partial_{\gamma} a_{0,{\beta_i\gamma}}^{(n-1)} \right]}_{(iv.ii)} \\
& \underbrace{-c_s^2 (\theta-1) \partial_{\alpha_n} a_{0,{\alpha_1..\alpha_{n-1}}}^{(n-1)}-c_s^2 a_{0,{\alpha_1..\alpha_{n-1}}}^{(n-1)} \partial_{\alpha_n} \theta}_{(iv.iii)},
\end{align*}
\begin{equation*}
(v) = \underbrace{-c_s^4 (\theta-1) \sum_{i=1}^{n-1} \sum_{j=1,j\ne i}^{n-1} \delta_{\alpha_j \alpha_n} \partial_{\alpha_i} a_{0,{\beta_{ij}}}^{(n-3)}}_{(v.i)} \underbrace{-c_s^4 \sum_{i=1}^{n-1} \sum_{j=1,j\ne i}^{n-1} \delta_{\alpha_j \alpha_n} a_{0,{\beta_{ij}}}^{(n-3)} \partial_{\alpha_i} \theta  }_{(v.ii)}.
\end{equation*}
Once again, one can replace $\partial_t a_{0,{\beta_i}}^{(n-2)}$ using~(\ref{eq:LBK_ProjectedOnH}),
$$ (iv.ii)+(v.i) = \frac{1}{\tau} c_s^2 (\theta-1) \sum_{i=1}^{n-1} \delta_{\alpha_i \alpha_n} a_{1,{\beta_i}}^{(n-2)}. $$
Moreover, since $\partial_{\alpha_n} p = \rho c_s^2 \partial_{\alpha_n} \theta + c_s^2 \theta \partial_{\alpha_n} \rho$ we can rearrange several terms as follows,
$$ (ii) + (iii) + (iv.iii) = a_{0,{\alpha_1..\alpha_{n-1}}}^{(n-1)} \frac{c_s^2 \theta}{\rho} \partial_{\alpha_n}\rho -c_s^2\theta \sum_{i=1}^{n-1}a_{0,{\beta_i}}^{(n-2)} \partial_{\alpha_i} u_{\alpha_n} - c_s^2 \theta \partial_{\alpha_n} a_{0,{\alpha_1..\alpha_{n-1}}}^{(n-1)}. $$
By using~(\ref{eq:Recursivity_a0}), one gets
\begin{align*}
	(iv.i) &= -c_s^2 \sum_{i=1}^{n-1} \delta_{\alpha_i \alpha_n} \Bigg[ a_{0,{\beta_i}}^{(n-2)} \partial_t \theta + \Big( u_{\gamma}a_{0,{\beta_i}}^{(n-2)}  + c_s^2(\theta-1) \sum_{j=1,j\ne i}^{n-1} \delta_{\alpha_j \gamma} a_{0,{\beta_{ij}}}^{(n-3)} \Big) \partial_{\gamma} \theta \Bigg] \\
	& = -c_s^2 \sum_{i=1}^{n-1} \delta_{\alpha_i \alpha_n} a_{0,{\beta_i}}^{(n-2)}(\partial_t \theta + u_{\gamma} \partial_{\gamma} \theta)  - c_s^4 (\theta -1) \partial_{\gamma} \theta \sum_{i=1}^{n-1} \sum_{j=1,j\ne i}^{n-1} \delta_{\alpha_i \alpha_n} \delta_{\alpha_j \gamma} a_{0,{\beta_{ij}}}^{(n-3)}.
\end{align*}
The last time derivative $\partial_t \theta$ is removed thanks to~(\ref{eq:Euler_temperature}). Further simplifications, using the implicit summation on $\gamma$ in the second term, lead to
\begin{align*}
(iv.i)+(v.ii)=\frac{2c_s^2 \theta}{D} \partial_{\gamma}u_{\gamma} \sum_{i=1}^{n-1} \delta_{\alpha_i \alpha_n} a_{0,{\beta_i}}^{(n-2)} - c_s^4 \theta \sum_{i=1}^{n-1} \sum_{j=1,j\ne i}^{n-1} \delta_{\alpha_i \alpha_n} a_{0,{\beta_{ij}}}^{(n-3)} \partial_{\alpha_j} \theta.
\end{align*}
Eventually, regrouping all terms leads to
\begin{align}
\label{eq:recursivity_tmp1}
\forall n \ge 3,\quad u_{\alpha_n} a_{1,{\alpha_1..\alpha_{n-1}}}^{(n-1)} &=a_{1,{\alpha_1..\alpha_n}}^{(n)} - c_s^2 (\theta-1) \sum_{i=1}^{n-1} \delta_{\alpha_i \alpha_n} a_{1,{\beta_i}}^{(n-2)} + \tau c_s^2 \theta \sum_{i=1}^{n-1} a_{0,{\beta_i}}^{(n-2)} \left( \partial_{\alpha_i} u_{\alpha_n} - \frac{2}{D} \delta_{\alpha_i \alpha_n} \partial_{\gamma} u_{\gamma} \right) \nonumber\\
&+ \tau c_s^4 \theta \sum_{i=1}^{n-1} \sum_{j=1, j\ne i}^{n-1} \delta_{\alpha_i \alpha_n} a_{0,{\beta_{ij}}}^{(n-3)}\partial_{\alpha_j} \theta - \tau c_s^2 \frac{\theta}{\rho} a_{0,{\alpha_1..\alpha_{n-1}}}^{(n-1)} \partial_{\alpha_n} \rho + \tau c_s^2 \theta \partial_{\alpha_n} a_{0,{\alpha_1..\alpha_{n-1}}}^{(n-1)}.\\
\nonumber
\end{align}
\underline{STEP 3: Dispose of $\partial_{\alpha_n} a_{0,{\alpha_1..\alpha_{n-1}}}^{(n-1)}$ \emph{via} the definition of the Maxwell-Boltzmann equilibrium state $f^{(0)}$.}\\\\
By the chain rule,
\begin{align*}
\forall n \ge 3,\quad \partial_{\alpha_n} a_{0,{\alpha_1..\alpha_{n-1}}}^{(n-1)} &=\int \partial_{\alpha_n} f^{(0)}(\boldsymbol{\xi}) \mathcal{H}_{\alpha_1..\alpha_{n-1}}^{(n-1)} \text{d} \boldsymbol{\xi} \\
&= \int \left( \partial_\rho f^{(0)} \partial_{\alpha_n} \rho + \partial_{u_{\gamma}} f^{(0)} \partial_{\alpha_n} u_{\gamma} + \partial_\theta f^{(0)} \partial_{\alpha_n} \theta \right) \mathcal{H}_{\alpha_1..\alpha_{n-1}}^{(n-1)} \text{d} \boldsymbol{\xi},
\end{align*}
and since $ f^{(0)} = \frac{\rho}{(2\pi c_s^2 \theta)^{D/2}}\exp\left[-\frac{(\boldsymbol{\xi-u})^2}{2 c_s^2 \theta}\right]$, the following derivatives can be obtained:
\begin{equation}
\left\{
\begin{array}{l @{\: = \:} l}
\partial_\rho f^{(0)} & \dfrac{1}{\rho} f^{(0)},\vspace*{0.2cm}\\ 
\partial_{u_{\gamma}} f^{(0)} & \dfrac{\xi_{\gamma} - u_{\gamma}}{c_s^2 \theta} f^{(0)},\vspace*{0.2cm}\\
\partial_\theta f^{(0)} & \left[ -\dfrac{D}{2 \theta} + \dfrac{1}{2c_s^2 \theta^2}(\xi_{\gamma} - u_{\gamma})^2 \right] f^{(0)}.
\end{array}
\right.
\end{equation}
Hence
\begin{align*}
\partial_{\alpha_n} a_{0,{\alpha_1..\alpha_{n-1}}}^{(n-1)} &= \frac{1}{\rho} a_{0,{\alpha_1..\alpha_{n-1}}}^{(n-1)} \partial_{\alpha_n} \rho + \frac{1}{c_s^2 \theta} \partial_{\alpha_n}u_{\gamma} \int (\xi_{\gamma} - u_{\gamma}) \mathcal{H}_{\alpha_1..\alpha_{n-1}}^{(n-1)} f^{(0)}(\boldsymbol{\xi}) \text{d}\boldsymbol{\xi}\\
 &- \frac{D}{2 \theta} a_{0,{\alpha_1..\alpha_{n-1}}}^{(n-1)} \partial_{\alpha_n} \theta + \frac{1}{2c_s^2 \theta^2} \partial_{\alpha_n} \theta \int (\xi_{\gamma}-u_{\gamma})^2 \mathcal{H}_{\alpha_1..\alpha_{n-1}}^{(n-1)}f^{(0)}(\boldsymbol{\xi}) \text{d} \boldsymbol{\xi},
\end{align*}
and by integration by parts, one can show that
\begin{equation}
\begin{array}{l @{\: = \:} l}
\displaystyle{\int (\xi_{\gamma} - u_{\gamma}) \mathcal{H}_{\alpha_1..\alpha_{n-1}}^{(n-1)} f^{(0)}(\boldsymbol{\xi}) \text{d}\boldsymbol{\xi}} & \theta c_s^2 \displaystyle{\sum_{i=1}^{n-1} \delta_{\alpha_i \gamma}a_{0,{\beta_i}}^{(n-2)}},\vspace*{0.1cm}\\
\displaystyle{\int (\xi_{\gamma}-u_{\gamma})^2 \mathcal{H}_{\alpha_1..\alpha_{n-1}}^{(n-1)}f^{(0)}(\boldsymbol{\xi}) \text{d} \boldsymbol{\xi}} & D c_s^2 \theta a_{0,{\alpha_1..\alpha_{n-1}}}^{(n-1)} + \theta^2 c_s^4 \displaystyle{\sum_{i=1}^{n-1} \sum_{j=1, j\ne i}^{n-1} \delta_{\alpha_i \alpha_j} a_{0,{\beta_{ij}}}^{(n-3)}}.
\end{array}
\end{equation}
Then, removing the implicit summation on $\gamma$ leads to
\begin{align*}
\forall n \ge 3,\quad \partial_{\alpha_n} a_{0,{\alpha_1..\alpha_{n-1}}}^{(n-1)} &= \frac{1}{\rho} a_{0,{\alpha_1..\alpha_{n-1}}}^{(n-1)} \partial_{\alpha_n} \rho + \sum_{i=1}^{n-1} a_{0,{\beta_i}}^{(n-2)} \partial_{\alpha_n}u_{\alpha_i} + \frac{c_s^2}{2} \sum_{i=1}^{n-1} \sum_{j=1}^{n-1} \delta_{\alpha_i \alpha_j}a_{0,{\beta_{ij}}}^{(n-3)} \partial_{\alpha_n}\theta.
\end{align*}
Injecting this expression in~(\ref{eq:recursivity_tmp1}) gives
\begin{align}
\forall n\geq 3,\quad a_{1,{\alpha_1..\alpha_n}}^{(n)} &= u_{\alpha_n} a_{1,{\alpha_1..\alpha_{n-1}}}^{(n-1)} + c_s^2 (\theta-1) \sum_{i=1}^{n-1} \delta_{\alpha_i \alpha_n} a_{1,{\beta_i}}^{(n-2)}
- \tau c_s^2 \theta \sum_{i=1}^{n-1} a_{0,{\beta_i}}^{(n-2)} \left( \partial_{\alpha_i} u_{\alpha_n} + \partial_{\alpha_n}u_{\alpha_i} - \frac{2}{D} \delta_{\alpha_i \alpha_n} \partial_{\gamma}u_{\gamma} \right) \nonumber\\
&- \tau c_s^4 \theta \sum_{i=1}^{n-1} \sum_{j>i}^{n-1} \left( \delta_{\alpha_i \alpha_n} \partial_{\alpha_j} \theta + \delta_{\alpha_j \alpha_n} \partial_{\alpha_i} \theta + \delta_{\alpha_i \alpha_j} \partial_{\alpha_n} \theta \right) a_{0,{\beta_{ij}}}^{(n-3)}.
\label{eq:recursivity_tmp2}
\end{align}
Finally, knowing that
\begin{equation*}
\hspace*{-0.2cm}\begin{array}{c}
a_{1,{\alpha \beta}}^{(2)} = -\tau \rho c_s^2 \theta \left(\partial_{\alpha}u_\beta + \partial_\beta u_\alpha - \dfrac{2}{D} \delta_{\alpha \beta} \partial_{\gamma} u_\gamma  \right) \equiv \Pi_{\alpha\beta}^{(1)}\approx \displaystyle{\sum_i}\xi_{\alpha}\xi_{\beta}\left(f_i-f^{(0)}_i\right),\vspace{0.1cm}\\
a_{1,{\alpha \beta \gamma}}^{(3)} = u_\alpha a_{1,{\beta \gamma}}^{(2)} + u_\beta a_{1,{\alpha \gamma}}^{(2)} + u_\gamma a_{1,{\alpha \beta}}^{(2)}- \tau \rho\theta c_s^4 \left(\delta_{\alpha \beta} \partial_\gamma \theta + \delta_{\alpha \gamma} \partial_{\beta} \theta + \delta_{\beta \gamma} \partial_\alpha \theta \right) \equiv Q_{\alpha\beta\gamma}^{(1)}\approx \displaystyle{\sum_i}\xi_{\alpha}\xi_{\beta}\xi_{\gamma}\left(f_i-f^{(0)}_i\right),
\end{array}
\end{equation*}
leads to~(\ref{eq:RecursiveRelation_ThermalCase_proof}), which is valid for $n \ge 4$.
\begin{flushright} \ding{111} \end{flushright}

\noindent\underline{Particularity of the isothermal case} \newline

The same kind of recursive relation can be obtained in the isothermal case, after noticing that this assumption has two important consequences in the previous derivation:
\begin{itemize}
	\item the temperature is constant ($\theta=1$),
	\item the term $-\dfrac{2}{D}\delta_{\alpha_i \alpha_n} \partial_\gamma u_\gamma$ disappears in (\ref{eq:recursivity_tmp2}), leading to a non-zero bulk viscosity $\mu_b = \dfrac{2}{D}\mu$~\cite{DELLAR_PRE_64_2001}.
\end{itemize}
Thus equation (\ref{eq:recursivity_tmp2}) becomes
\begin{equation}
	\forall n \ge 3,\ a_{1, \alpha_1..\alpha_n}^{(n)} = u_{\alpha_n} a_{1, \alpha_1..\alpha_{n-1}}^{(n-1)} - \tau c_s^2 \sum_{i=1}^{n-1} a_{0,\beta_i}^{(n-2)} \left( \partial_{\alpha_i} u_{\alpha_n} + \partial_{\alpha_n} u_{\alpha_i} \right),
\end{equation}
and knowing that
$$ a_{1, \alpha \beta}^{(2)} = - \tau \rho c_s^2 (\partial_\alpha u _\beta + \partial_\beta u_\alpha), $$
Malaspinas' recursive formula~\cite{MALASPINAS_ARXIV_2015} can be recovered:
\begin{equation}
	a_{1,{\alpha_1..\alpha_n}}^{(n)}=u_{\alpha_{n}} a_{1,{\alpha_1..\alpha_{n-1}}}^{(n-1)} + \frac{1}{\rho} \sum_{i=1}^{n-1} a_{0, \beta_i} ^{(n-2)} a_{1,\alpha_i \alpha_n}^{(2)}
\end{equation}

\section{\label{sec:appendixB}Recursive formula over equilibrium coefficients}
\noindent The aim of this section is to prove the following relation
\begin{equation}
	\label{eq:RecursivityEquilibrium}
	\forall n \ge 2,\quad a_{0,\alpha_1..\alpha_n}^{(n)} = u_{\alpha_n} a_{0,\alpha_1..\alpha_{n-1}}^{(n-1)} + (\theta-1)c_s^2 \sum_{i=1}^{n-1} \delta_{\alpha_i \alpha_n} a_{0,\beta_i}^{(n-2)},
\end{equation}
which simplifies in 
$$\forall n \ge 2,\quad a_{0,\alpha_1..\alpha_n}^{(n)} = u_{\alpha_n} a_{0,\alpha_1..\alpha_{n-1}}^{(n-1)} $$
for the isothermal case ($\theta = 1$). To this end, the Rodrigues relation over Hermite polynomials will be used:
\begin{equation}
	\label{eq:Rodrigues}
	\forall n \ge 2,\quad \xi_{\alpha_{n}} \mathcal{H}_{\alpha_1..\alpha_{n-1}}^{(n-1)} = \mathcal{H}_{\alpha_1..\alpha_{n}}^{(n)} + c_s^2 \sum_{i=1}^{n-1} \delta_{\alpha_i \alpha_{n}} \mathcal{H}_{\beta_i}^{(n-2)}.
\end{equation}
Using this relation, one gets, for $n \ge 2$:
\begin{align*}
a_{0,\alpha_1..\alpha_n}^{(n)} &= \int \mathcal{H}_{\alpha_1..\alpha_n}^{(n)}(\boldsymbol{\xi}) f^{(eq)}(\boldsymbol{\xi}) \mathrm{d} \boldsymbol{\xi} = \int \xi_{\alpha_n} \mathcal{H}_{\alpha_1..\alpha_{n-1}}^{(n-1)} f^{(eq)}(\boldsymbol{\xi}) \mathrm{d} \boldsymbol{\xi}-c_s^2 \sum_{i=1}^{n-1} \delta_{\alpha_i \alpha_n} \int \mathcal{H}_{\beta_i}^{(n-2)}f^{(eq)}(\boldsymbol{\xi}) \mathrm{d} \boldsymbol{\xi} \\
& = \underbrace{\int c_{\alpha_n} \mathcal{H}_{\alpha_1..\alpha_{n-1}}^{(n-1)}f^{(eq)}(\boldsymbol{\xi}) \mathrm{d} \boldsymbol{\xi}}_{I} + u_{\alpha_n} a_{0,\alpha_1..\alpha_{n-1}}^{(n-1)}-c_s^2 \sum_{i=1}^{n-1} \delta_{\alpha_i \alpha_n} a_{0,\beta_i}^{(n-2)},
\end{align*}
where $\boldsymbol{c}=\boldsymbol{\xi}-\boldsymbol{u}$. By integration by parts, the first term gives
$$ I = \theta c_s^2 \int \partial_{\xi_{\alpha_n}} \mathcal{H}_{\alpha_1..\alpha_{n-1}}^{(n-1)}f^{(eq)}(\boldsymbol{\xi}) \mathrm{d} \boldsymbol{\xi}. $$
Finally, using the fact that
$$ \partial_{\xi_{\alpha_n}} \mathcal{H}_{\alpha_1..\alpha_{n-1}}^{(n-1)} = \sum_{i=1}^{n-1} \delta_{\alpha_i \alpha_n} \mathcal{H}_{\beta_i}^{(n-2)},$$
one gets
$$I = \theta c_s^2 \sum_{i=1}^{n-1} \delta_{\alpha_i \alpha_n} a_{0,\beta_i}^{(n-2)}, $$
so that
$$\forall n \ge 2,\quad a_{0,\alpha_1..\alpha_n}^{(n)} = u_{\alpha_n} a_{0,\alpha_1..\alpha_{n-1}}^{(n-1)} + (\theta-1)c_s^2 \sum_{i=1}^{n-1} \delta_{\alpha_i \alpha_n} a_{0,\beta_i}^{(n-2)}.$$
\begin{flushright} \ding{111} \end{flushright}

\section{\label{sec:appendixC}Implementation details}
This appendix summarizes all the necessary material to properly implement the recursive regularization procedure, in both 2D and 3D, for some common and high-order lattice structures. As a reminder, the purpose of this step is to rebuilt the discrete VDF $f_i$ knowing the macroscopic properties of the flow (density $\rho$, velocity $\boldsymbol{u}$ and temperature $\theta$) and their gradients:
\begin{equation}
f_i^{reg} = f_i^{(0),reg} + f_i^{(1),reg} = \displaystyle{\sum_n\dfrac{1}{n!c_s^{2n}}\mathcal{H}^{(n)}_{i,\boldsymbol\alpha}\left(a^{(n)}_{0,\boldsymbol\alpha}+a^{(n)}_{1,\boldsymbol\alpha}\right)}.
 \end{equation} 
\\ 
\paragraph*{\textbf{Hermite tensors:}}
The definition of Hermite tensors up to the fourth order (sufficient for the recovery of Navier-Stokes' equations) is
\begin{equation}
\begin{array}{l @{\: = \:} l}
\mathcal{H}_i^{(0)} & 1, \vspace{0.1cm}\\ 
\mathcal{H}^{(1)}_{i,\alpha} & \xi_{i,\alpha}, \vspace{0.1cm}\\ 
\mathcal{H}^{(2)}_{i,\alpha\beta} & \xi_{i,\alpha\beta} - c_s^2\delta_{\alpha\beta}, \vspace{0.1cm}\\ 
\mathcal{H}^{(3)}_{i,\alpha\beta\gamma} & \xi_{i,\alpha\beta\gamma} - c_s^2\left(\xi_{i,\alpha}\delta_{\beta\gamma} + \xi_{i,\beta}\delta_{\alpha\gamma} + \xi_{i,\gamma}\delta_{\alpha\beta}\right), \vspace{0.1cm}\\ 
\mathcal{H}^{(4)}_{i,\alpha\beta\gamma\delta} & \xi_{i,\alpha\beta\gamma\delta}  - c_s^2\left(\xi_{i,\alpha\beta}\delta_{\gamma\delta}+\xi_{i,\alpha\gamma}\delta_{\beta\delta}+\xi_{i,\alpha\delta}\delta_{\beta\gamma}
+ \xi_{i,\beta\gamma}\delta_{\alpha\delta}+\xi_{i,\beta\delta}\delta_{\alpha\gamma}+\xi_{i,\gamma\delta}\delta_{\alpha\beta}\right) + c_s^4\left(\delta_{\alpha\beta}\delta_{\gamma\delta}+\delta_{\alpha\gamma}\delta_{\beta\delta} + \delta_{\alpha\delta}\delta_{\beta\gamma}\right),
\end{array}
\end{equation}
where the following mathematical notations $\xi_{i,\alpha\beta}\equiv\xi_{i,\alpha}\xi_{i,\beta}$, $\xi_{i,\alpha\beta\gamma}\equiv\xi_{i,\alpha}\xi_{i,\beta}\xi_{i,\gamma}$, $\xi_{i,\alpha\beta\gamma\delta}\equiv\xi_{i,\alpha}\xi_{i,\beta}\xi_{i,\gamma}\xi_{i,\delta}$ is used for the sake of compacity.

\noindent In 2D, $(\alpha,\beta,\gamma,\delta)\in \{x,y\}^4$ and thus only two different values can be chosen for each index. This simplifies the definition of Hermite polynomials into:
\begin{equation}
\begin{array}{l @{\: = \:} l}
\mathcal{H}^{(0)}_i & 1, \vspace{0.1cm}\\ 
\mathcal{H}^{(1)}_{i,x} & \xi_{i,x}, \vspace{0.1cm}\\ 
\mathcal{H}^{(2)}_{i,xx} & \xi_{i,x}^2 - c_s^2, \quad\mathcal{H}^{(2)}_{i,xy} = \xi_{i,x}\xi_{i,y}, \vspace{0.1cm}\\ 
\mathcal{H}^{(3)}_{i,xxx} & \left(\xi_{i,x}^2 - 3c_s^2\right)\xi_{i,x}, \quad\mathcal{H}^{(3)}_{i,xxy} = \left(\xi_{i,x}^2 - c_s^2\right)\xi_{i,y}, \vspace{0.1cm}\\  
\mathcal{H}^{(4)}_{i,xxxx} & \xi_{i,x}^4  - 6 c_s^2 \xi_{i,x}^2 + 3 c_s^4, \quad\mathcal{H}^{(4)}_{i,xxxy} = \left(\xi_{i,x}^2  - 3c_s^2\right) \xi_{i,x}\xi_{i,y}, \quad\mathcal{H}^{(4)}_{i,xxyy} = (\xi_{i,x}^2-c_s^2)(\xi_{i,y}^2 -c_s^2).
\end{array}
\end{equation}
 
\noindent In 3D, $(\alpha,\beta,\gamma,\delta)\in \{x,y,z\}^4$ and therefore one more value can be chosen for each index. This adds the following Hermite polynomials:
\begin{equation}
\begin{array}{l @{\: = \:} l}
\mathcal{H}^{(3)}_{i,xyz} & \xi_{i,x}\xi_{i,y}\xi_{i,z}, \vspace{0.1cm}\\  
\mathcal{H}^{(4)}_{i,xxyz} & \xi_{i,x}^2\xi_{i,y}\xi_{i,z} -c_s^2\xi_{i,y}\xi_{i,z}+c_s^4.
\end{array}
\end{equation}
Due to symmetry properties of Hermite tensors ($\mathcal{H}^{(1)}_{i,y}$, $\mathcal{H}^{(1)}_{i,z}$, $\mathcal{H}^{(2)}_{i,yy}$, $\mathcal{H}^{(2)}_{i,zz}$,$\mathcal{H}^{(2)}_{i,xz}$,$\mathcal{H}^{(2)}_{i,yz}$, $\mathcal{H}^{(3)}_{i,yyy}$, $\mathcal{H}^{(3)}_{i,zzz}$, $\mathcal{H}^{(3)}_{i,xxz}$, $\mathcal{H}^{(3)}_{i,yyx}$, $\mathcal{H}^{(3)}_{i,yyz}$, $\mathcal{H}^{(4)}_{i,yyyy}$, $\mathcal{H}^{(4)}_{i,zzzz}$, $\mathcal{H}^{(4)}_{i,xxxz}$, $\mathcal{H}^{(4)}_{i,yyyx}$, $\mathcal{H}^{(4)}_{i,yyyz}$, $\mathcal{H}^{(4)}_{i,xxzz}$, $\mathcal{H}^{(4)}_{i,yyzz}$, $\mathcal{H}^{(4)}_{i,yyxz}$, $\mathcal{H}^{(4)}_{i,zzxy}$) can be obtained changing $x$ by $y$ or $z$ in the above formulas. By definition, the very same property is also available for $a_{0,\boldsymbol{\alpha}}^{(n)}$ and $a_{1,\boldsymbol{\alpha}}^{(n)}$.
\\
\paragraph*{\textbf{Hermite coefficients at equilibrium:}}
The recursive definition of Hermite coefficients at equilibrium is first recalled~(\ref{eq:RecursivityEquilibrium}). Up to the fourth order, they read
\begin{equation}
\begin{array}{l @{\: = \:} l}
a_{0}^{(0)}&\rho, \vspace*{0.1cm}\\
a_{0,\alpha}^{(1)}&u_{\alpha} a_{0}^{(0)}, \vspace*{0.1cm}\\	
a_{0,\alpha\beta}^{(2)}&u_{\beta}a_{0,\alpha}^{(1)} + (\theta -1)c_s^2\delta_{\alpha\beta} a_{0}^{(0)}, \vspace*{0.1cm}\\	
a_{0,\alpha\beta\gamma}^{(3)}&u_{\gamma}a_{0,\alpha\beta}^{(2)} +(\theta -1)c_s^2 \left(\delta_{\alpha\gamma}a_{0,\beta}^{(1)} + \delta_{\beta\gamma}a_{0,\alpha}^{(1)}\right) , \vspace*{0.1cm}\\	
a_{0,\alpha\beta\gamma\delta}^{(4)}&u_{\delta}a_{0,\alpha\beta\gamma}^{(3)} +(\theta -1)c_s^2 \left(\delta_{\alpha\delta}a_{0,\beta\gamma}^{(2)} + \delta_{\beta\delta}a_{0,\alpha\gamma}^{(2)} + \delta_{\gamma\delta}a_{0,\alpha\beta}^{(2)}\right).
\end{array}
\end{equation}
This becomes in 2D,
\begin{equation}
\begin{array}{l @{\: = \:} l}
a_{0}^{(0)}&\rho, \vspace*{0.1cm}\\
a_{0,x}^{(1)}&u_{x} a_{0}^{(0)}, \vspace*{0.1cm}\\	
a_{0,xx}^{(2)}&u_{x}a_{0,x}^{(1)} + (\theta -1)c_s^2 a_{0}^{(0)}, \quad a_{0,xy}^{(2)} = u_{y}a_{0,x}^{(1)}, \vspace*{0.1cm}\\	
a_{0,xxx}^{(3)}&u_{x}a_{0,xx}^{(2)} +3(\theta -1)c_s^2 a_{0,x}^{(1)}, \quad a_{0,xxy}^{(3)}=u_{y}a_{0,xx}^{(2)}, \vspace*{0.1cm}\\	
a_{0,xxxx}^{(4)}&u_{x}a_{0,xxx}^{(3)} +3(\theta -1)c_s^2 a_{0,xx}^{(2)}, \quad a_{0,xxxy}^{(4)}=u_{y}a_{0,xxx}^{(3)}, \quad a_{0,xxyy}^{(4)}=u_{y}a_{0,xxy}^{(3)} +(\theta -1)c_s^2 a_{0,xx}^{(2)}.	
\end{array}
\end{equation}

\noindent And the 3D extension adds the following coefficients,
\begin{equation}
\begin{array}{l @{\: = \:} l}
a_{0,xyz}^{(3)}&u_{z}a_{0,xy}^{(2)}, \vspace*{0.1cm}\\	
a_{0,xxyz}^{(4)}&u_{z}a_{0,xxy}^{(3)}.	
\end{array}
\end{equation}
\\
\paragraph*{\textbf{First-order off-equilbrium Hermite coefficients:}}
The recursive definition of first-order (with respect to the Knudsen number) off-equilibrium Hermite coefficients (Eq.~(\ref{eq:RecursiveRelation_ThermalCase})), up to the fourth order (with respect to the VDF moments), reads
\begin{align}
a_{1}^{(0)} &= 0, \nonumber\\[0.1cm]
a_{1,\alpha}^{(1)} &= 0, \nonumber\\[0.1cm]
a_{1,\alpha\beta}^{(2)} &= \Pi^{(1)}_{\alpha\beta} = -\mu c_s^2\theta\big[S_{\alpha\beta}  - \left(\textstyle{\frac{2}{D}}\partial_{\gamma}u_{\gamma}\right)\delta_{\alpha\beta}\big], \nonumber\\[0.1cm]
a_{1,\alpha\beta\gamma}^{(3)} &= Q^{(1)}_{\alpha\beta\gamma} = \left(u_{\alpha}a^{(2)}_{1,\beta\gamma} + u_{\beta}a^{(2)}_{1,\alpha\gamma}+u_{\gamma}a^{(2)}_{1,\alpha\beta}\right) -\mu c_s^2 \left(\delta_{\alpha\beta}\partial_{\gamma}\theta+\delta_{\alpha\gamma}\partial_{\beta}\theta+\delta_{\beta\gamma}\partial_{\alpha}\theta\right),\\[0.1cm]
a_{1,\alpha\beta\gamma\delta}^{(4)} &= \left(u_{\alpha}a_{1,\beta\gamma\delta}^{(3)}+u_{\beta}a_{1,\alpha\gamma\delta}^{(3)} + u_{\gamma}a_{1,\alpha\beta\delta}^{(3)}+u_{\delta}a_{1,\alpha\beta\gamma}^{(3)}\right) +\left[c_s^2(\theta -1)\delta_{\alpha\beta}
- u_{\alpha}u_{\beta}\right]a^{(2)}_{1,\gamma\delta} \nonumber\\[0.1cm]
&+ \left[c_s^2(\theta -1)\delta_{\alpha\gamma} - u_{\alpha}u_{\gamma}\right]a^{(2)}_{1,\beta\delta} + \left[c_s^2(\theta -1)\delta_{\alpha\delta} - u_{\alpha}u_{\delta}\right]a^{(2)}_{1,\beta\gamma}+ \left[c_s^2(\theta -1)\delta_{\beta\gamma} - u_{\beta}u_{\gamma}\right]a^{(2)}_{1,\alpha\delta} \nonumber\\[0.1cm] 
&+ \left[c_s^2(\theta -1)\delta_{\beta\delta} - u_{\beta}u_{\delta}\right]a^{(2)}_{1,\alpha\gamma} + \left[c_s^2(\theta -1)\delta_{\gamma\delta} - u_{\gamma}u_{\delta}\right]a^{(2)}_{1,\alpha\beta}, \nonumber
\end{align}
with $S_{\alpha\beta} = \partial_{\alpha}u_{\beta} + \partial_{\beta}u_{\alpha}$, and $\mu=\rho c_s^2\theta \tau$ is the dynamic viscosity.\\

\noindent For the 2D case, this simplifies into,
\begin{align}
a_{1}^{(0)} &= 0, \nonumber\\[0.1cm]
a_{1,x}^{(1)} &= 0, \nonumber\\[0.1cm]
a_{1,xx}^{(2)} &= \Pi^{(1)}_{xx} = -\mu c_s^2\theta\big[S_{xx}  - \textstyle{\frac{2}{D}}\partial_{\gamma}u_{\gamma}\big], \quad a_{1,xy}^{(2)} = \Pi^{(1)}_{xy} = -\tau\rho c_s^2\theta S_{xy} \nonumber\\[0.1cm]
a_{1,xxx}^{(3)} &= Q^{(1)}_{xxx} = 3 \left(u_{x}a^{(2)}_{1,xx}  -\mu c_s^2 \partial_{x}\theta\right),
\quad a_{1,xxy}^{(3)} = Q^{(1)}_{xxy} = \left(2 u_{x}a^{(2)}_{1,xy} + u_{y}a^{(2)}_{1,xx}\right) -\mu c_s^2 \partial_{y}\theta,\label{eq:a1Rec}\\[0.1cm]
a_{1,xxxx}^{(4)} &= 4 u_{x}a_{1,xxx}^{(3)} +6\left[c_s^2(\theta -1)
- u_{x}^2\right]a^{(2)}_{1,xx}, \nonumber\\[0.1cm] 
a_{1,xxxy}^{(4)} &= \left(3 u_{x}a_{1,xxy}^{(3)}+u_{y}a_{1,xxx}^{(3)} \right) +3 \left[c_s^2(\theta -1)
- u_{x}^2\right]a^{(2)}_{1,xy} - 3u_{x}u_{y}a^{(2)}_{1,xx}, \nonumber\\[0.1cm] 
a_{1,xxyy}^{(4)} &= 2\left(u_{x}a_{1,yyx}^{(3)}+u_{y}a_{1,xxy}^{(3)}\right) +\left[c_s^2(\theta -1)
- u_{x}^2\right]a^{(2)}_{1,yy} +\left[c_s^2(\theta -1) - u_{y}^2\right]a^{(2)}_{1,xx} -4u_{x}u_{y} a^{(2)}_{1,xy}. \nonumber
\end{align}
And for the 3D extension, more coefficients need to be taken into account,
\begin{equation}
\begin{array}{l @{\: = \:} l}
a_{1,xyz}^{(3)} & Q^{(1)}_{xyz} = u_{x}a^{(2)}_{1,yz} + u_{y}a^{(2)}_{1,xz}+u_{z}a^{(2)}_{1,xy},\\[0.1cm]
a_{1,xxyz}^{(4)} & \left(2u_{x}a_{1,xyz}^{(3)}+u_{y}a_{1,xxz}^{(3)} + u_{z}a_{1,xxy}^{(3)}\right) +\left[c_s^2(\theta -1)
- u_{x}^2\right]a^{(2)}_{1,yz} -  2u_{x}\left(u_{y}a^{(2)}_{1,xz} + u_{z}a^{(2)}_{1,xy} \right) - u_{y}u_{z}a^{(2)}_{1,xx}.
\end{array}
\end{equation}
Here attention must be paid to the way $a_{1,\alpha\beta}^{(2)}$ and $a_{1,\alpha\beta\gamma}^{(3)}$ are computed. One can use finite differences to evaluate the velocity and temperature gradients, but it will degrade the accuracy of the algorithm. Instead, we propose the following idea for LBM allowing the preservation of moments of the VDF up to the fourth order ($D2V37$ and $D3Q103$):
\begin{enumerate}
\item Compute $a_{1,\alpha\beta}^{(2)}$ using the definition of the second-order moment of the VDF,
\begin{equation}
a_{1,\alpha\beta}^{(2)} \approx \displaystyle{\sum_i}\mathcal{H}^{(2)}_{\alpha\beta}\Big(f_i-f_i^{(0)}\Big).
\label{eq:a12Proj}
\end{equation}
This is justified by the fact that the error introduced by the approximation $f_i = \sum_n f_i^{(n)} \approx f_i^{(0)} + f_i^{(1)}$, with $f_i^{(n)} ~\sim \mathcal{O}\left(\epsilon^n\right)$, should be small enough when the LBM allows to conserve moments of the VDF up to the fourth order.
\item Compute $a_{1,\alpha\beta\gamma}^{(3)}$ using the definition of the third-order moment of the VDF, 
\begin{equation}
a_{1,\alpha\beta\gamma}^{(3)} \approx \displaystyle{\sum_i}\mathcal{H}^{(3)}_{\alpha\beta\gamma}\Big(f_i-f_i^{(0)}\Big).
\end{equation}
\item Average the contributions of all $a_{1,\alpha\beta\gamma}^{(3)}$ to the temperature gradients,
\begin{align}
\mu c_s^2 \partial_{x}\theta &\overset{2D}{=} 
\dfrac{1}{2}\left[\left(u_{x}a_{1,xx}^{(2)}-a_{1,xxx}^{(3)}/3\right) + \left(2u_{y}a_{1,xy}^{(2)} + u_{x}a_{1,yy}^{(2)}-a_{1,yyx}^{(3)}\right)\right]\vspace*{0.1cm}\\
&\overset{3D}{=} \dfrac{1}{3}\left[\left(u_{x}a_{1,xx}^{(2)}-a_{1,xxx}^{(3)}/3\right) + \left(2u_{y}a_{1,xy}^{(2)} + u_{x}a_{1,yy}^{(2)}-a_{1,yyx}^{(3)}\right) + \left(2u_{z}a_{1,xz}^{(2)} + u_{x}a_{1,zz}^{(2)}-a_{1,zzx}^{(3)}\right)\right]
\label{eq:TempGradLBM}
\end{align}
\item Reconstruct $a_{1,\alpha\beta\gamma}^{(3)}$ using Eq.~(\ref{eq:a1Rec}) thanks to Eqs.~(\ref{eq:a12Proj}) \&~(\ref{eq:TempGradLBM}).\\  
\end{enumerate}

\paragraph*{\textbf{Regularized collision operator:}}
The regularization procedure of pre-collision VDFs can be interpreted as a particular collision step where
 \begin{equation}
 f_i^{coll} = f_i^{(0),reg} + \left(1-\dfrac{1}{\tau}\right)f_i^{(1),reg} = \omega_i \displaystyle{\sum_n\dfrac{A^{(n)}_{\boldsymbol{\alpha}}}{n!c_s^{2n}}\mathcal{H}^{(n)}_{i,\boldsymbol\alpha}\left(a^{(n)}_{0,\boldsymbol\alpha}+\left(1-\dfrac{1}{\tau_{n,\boldsymbol{\alpha}}}\right)a^{(n)}_{1,\boldsymbol\alpha}\right)},
 \label{eq:GenRegCollision}
 \end{equation}
with $A^{(n)}_{\boldsymbol{\alpha}}$ being the number of times each Hermite tensors appears in the expansion. As an example, $\mathcal{H}^{(3)}_{i,xxy}$ appears three times in the above development since $\mathcal{H}^{(3)}_{i,xxy}=\mathcal{H}^{(3)}_{i,xyx}=\mathcal{H}^{(3)}_{i,yxx}$. This property is taken into account imposing $A^{(3)}_{xxy} = 3$. And more generally,
\begin{align}
A^{(n)}_{\boldsymbol{\alpha}} &\overset{2D}{=}
\begin{pmatrix}
n \\
n_x
\end{pmatrix}=\dfrac{n!}{n_x!(n-n_x)!} =\dfrac{(n_x+n_y)!}{n_x!n_y!},\\[0.1cm]
&\overset{3D}{=}
\begin{pmatrix}
n \\
n_x
\end{pmatrix}
\begin{pmatrix}
(n-n_x) \\
n_y
\end{pmatrix} 
=\dfrac{(n_x+n_y+n_z)!}{n_x!n_y!n_z!}.  
\end{align}
It flows from Pascal's triangle and pyramid rules (also called binomial and trinomial expansions) using $n_x$, $n_y$ and $n_z$ the number of occurrences of $x$, $y$ and $z$ in $\boldsymbol{\alpha}=(\alpha_1,..,\alpha_n)$ respectively. It should be noted that Eq.~(\ref{eq:GenRegCollision}) is the most general form of the regularized collision step, which allows to decouple the relaxation process specific to each Hermite coefficients. A similar formula was previously given by Shan \& Chen~\cite{SHAN_IJMPC_18_2007} where all Hermite coefficients belonging to the same expansion order $n$ were related by the same relaxation time, \emph{i.e.}, $\tau_{n,\boldsymbol{\alpha}}\longrightarrow \tau_n$. The latter formulation allows to preserve isotropy properties of the LBM and is thus preferred.
\renewcommand{\arraystretch}{1.4}
\begin{table}[bp!]
\begin{tabular}{c @{\quad} c @{\quad\quad} c @{\quad} | @{\quad\quad} c @{\quad\quad} c @{\quad\quad} c}
\hline\hline
$\text{Group}$ & $\boldsymbol\xi_i$ & $p$ & $E_{2,5}^{9}$ & $E_{2,7}^{17}$ & $E_{2,9}^{37}$\\
\hline
$1$ & $(0,0)$ & $1$ & $4/9$ & $( 575+ 193\sqrt{193} )/8100$ & $0.23315066913235250228650$\\
\hline
$2$ & $(1,0)$ & $4$ & $1/9$ & $(3355-91\sqrt{193})/18000$ & $0.10730609154221900241246$\\
$3$ & $(1,1)$ & $4$ & $1/9$ & $(655+17\sqrt{193})/27000$ & $0.05766785988879488203006$\\
\hline
$4$ & $(2,0)$ & $4$ & $1/36$ &  & $0.01420821615845075026469$\\
$5$ & $(2,1)$ & $8$ &  &   & $0.00535304900051377523273$\\
$6$ & $(2,2)$ & $4$ &  & $(685-49\sqrt{193})/54000$ & $0.00101193759267357547541$\\
\hline
$7$ & $(3,0)$ & $4$ &  & $(1445-101\sqrt{193})/162000$ & $0.00024530102775771734547$\\
$8$ & $(3,1)$ & $8$ &  &  & $0.00028341425299419821740$\\
\hline
\multicolumn{3}{c |@{\quad\quad}}{$\:\:1/c_s$} & $\sqrt{3}$ & $\sqrt{5(25 + \sqrt{193})/72}$ & $1.19697977039307435897239$ \\
\hline\hline\\
\end{tabular}
\begin{tabular}{ c @{\:\:}|@{\quad} c c c c c c c c c c }
\hline\hline\\[-0.35cm]
Lattice & $\mathcal{H}^{(0)}$ & $\mathcal{H}_x^{(1)}$ & $\mathcal{H}_{xx}^{(2)}$ & $\mathcal{H}_{xy}^{(2)}$ & $\mathcal{H}_{xxx}^{(3)}$ & $\mathcal{H}_{xxy}^{(3)}$ & $\mathcal{H}_{xxxx}^{(4)}$ & $\mathcal{H}_{xxxy}^{(4)}$ & $\mathcal{H}_{xxyy}^{(4)}$  \\[0.1cm]
\hline\\[-0.3cm]
$E_{2,5}^9$ & \ding{109} & \ding{109} & \ding{109} & \ding{109} & \ding{55} & \ding{109} & \ding{55} & \ding{55} & \ding{109} \\[0.2cm]
$E_{2,7}^{17}$ & \ding{109} & \ding{109} & \ding{109} & \ding{109} & \ding{109} & \ding{109} & \ding{55} & \ding{55} & \ding{55} \\[0.2cm]
$E_{2,9}^{37}$ & \ding{109} & \ding{109} & \ding{109} & \ding{109} & \ding{109} & \ding{109} & \ding{109} & \ding{109} & \ding{109} \\[0.2cm]
\hline\hline
\end{tabular}
\caption{Description of some common standard and high-order two-dimensional lattice structures (top), and their associated Hermite tensor basis (bottom). For each lattice structure, the convention $E_{D,Q}^{V}$ is adopted to summarize all the characteristics of interest, namely, the number of discrete velocities $V$, the quadrature order $Q$, and the number of dimensions $D$~\cite{SHAN_JCS_17_2016}. Furthermore, $p$ stands for the number of discrete speeds of each velocity group, while their associated weights $\omega_i$ compose the right part of the table. The last row consists of the value of the normalization constant (the inverse of the lattice constant $c_s$) needed for the on-grid property of the lattice structure ($(\xi_{i,x},\xi_{i,y})\in \mathbb{Z}^2$). Regarding the Hermite tensors, they are classified into two categories: those belonging to the basis (\ding{109}) and those which do not (\ding{55}). Lattice structures data are compiled from~\cite{SUCCI_Book_2002,PHILIPPI_PRE_73_2006}.}
\label{tab:2DLattices}
\end{table}

\section{\label{sec:appendixD}Lattice structures and Related Hermite tensors}

\renewcommand{\arraystretch}{1.4}
\begin{table}[bp!]
\begin{tabular}{c @{\quad} c @{\quad\quad} c @{\quad} | @{\quad\quad} c @{\quad\quad}  c @{\quad\quad} c @{\quad\quad} c}
\hline\hline
$\text{Group}$ & $\boldsymbol\xi_i$ & $p$ & $E_{3,5}^{19}$  & $E_{3,5}^{27}$ & $E_{3,7}^{39}$ & $E_{3,9}^{103}$\\
\hline
$1$ & $(0,0,0)$ & $1$ & $1/3$ & $8/27$ & $1/12$ & $0.032633351764471159466$ \\
\hline
$2$ & $(1,0,0)$ & $6$ & $1/18$ & $2/27$ & $1/12$ & $0.097656833590334574221$\\
$3$ & $(1,1,0)$ & $12$ & $1/36$ & $1/54$ &  &  \\
$4$ & $(1,1,1)$ & $8$ &  & $1/216$ & $1/27$ & $0.028097750290257335627$ \\
\hline
$5$ & $(2,0,0)$ & $6$ & &  & $2/135$ & $0.001045259560430061466$ \\
$6$ & $(2,1,0)$ & $24$ & &  &  & $0.005705329016894815990$ \\
$7$ & $(2,2,0)$ & $12$ & &  & $1/432$ & $0.000611939269829747839$\\
$8$ & $(2,2,2)$ & $8$ & &  &  & $0.000155964159374283722$\\
\hline
$9$ & $(3,0,0)$ & $6$ & &  & $1/1620$ & $0.000284443251800055207$ \\
$10$ & $(3,1,1)$ & $24$ & &  &  & $0.000130698375985191585$ \\
$11$ & $(3,3,3)$ & $8$ & &  &  & $0.000001223194501323058$ \\
\hline
\multicolumn{3}{c |@{\quad\quad}}{$\,1/c_s$} & $\sqrt{3}$ & $\sqrt{3}$ & $\sqrt{3/2}$ & $1.19697977039307435897239$\\
\hline\hline\\
\end{tabular}
\begin{tabular}{ c @{\:\:}|@{\quad} c @{\quad} c @{\quad} c @{\quad} c @{\quad} c @{\quad} c @{\quad} c @{\quad} c @{\quad} c @{\quad} c @{\quad} c @{\quad} c @{\quad} c}
\hline\hline\\[-0.35cm]
Lattice & $\mathcal{H}^{(0)}$ & $\mathcal{H}_x^{(1)}$ & $\mathcal{H}_{xx}^{(2)}$ & $\mathcal{H}_{xy}^{(2)}$ & $\mathcal{H}_{xxx}^{(3)}$ & $\mathcal{H}_{xxy}^{(3)}$ & $\mathcal{H}_{xyz}^{(3)}$ & $\mathcal{H}_{xxxx}^{(4)}$ & $\mathcal{H}_{xxxy}^{(4)}$ & $\mathcal{H}_{xxyy}^{(4)}$  & $\mathcal{H}_{xxyz}^{(4)}$ & $\mathcal{H}_{xxyyz}^{(5)}$ & $\mathcal{H}_{xxyyzz}^{(6)}$\\[0.1cm]
\hline\\[-0.3cm]
$E_{3,5}^{19}$ & \ding{109} & \ding{109} & \ding{109} & \ding{109} & \ding{55} & \ding{109} & \ding{55} & \ding{55} & \ding{55} & \ding{55} & \ding{55} & \ding{55} & \ding{55} \\[0.2cm]
$E_{3,5}^{27}$ & \ding{109} & \ding{109} & \ding{109} & \ding{109} & \ding{55} & \ding{109} & \ding{109} & \ding{55} & \ding{55} & \ding{109} & \ding{55} & \ding{109} & \ding{109} \\[0.2cm]
$E_{3,7}^{39}$ & \ding{109} & \ding{109} & \ding{109} & \ding{109} & \ding{109} & \ding{109} & \ding{109} & \ding{55} & \ding{55} & \ding{55} & \ding{55} & \ding{55} & \ding{55}\\[0.2cm]
$E_{3,9}^{103}$ & \ding{109} & \ding{109} & \ding{109} & \ding{109} & \ding{109} & \ding{109} & \ding{109} & \ding{109} & \ding{109} & \ding{109} & \ding{109} & \ding{55} & \ding{55}\\[0.2cm]
\hline\hline
\end{tabular}
\caption{Description of some common standard and high-order three-dimensional lattice structures (top), and their associated Hermite tensor basis (bottom). For each lattice structure, the convention $E_{D,Q}^{V}$ is adopted to summarize all the characteristics of interest, namely, the number of discrete velocities $V$, the quadrature order $Q$, and the number of dimensions $D$~\cite{SHAN_JCS_17_2016}. Furthermore, $p$ stands for the number of discrete speeds of each velocity group, while their associated weights $\omega_i$ compose the right part of the table. The last row consists of the value of the normalization constant (the inverse of the lattice constant $c_s$) needed for the on-grid property of the lattice structure ($(\xi_{i,x},\xi_{i,y},\xi_{i,z})\in \mathbb{Z}^3$). Regarding the Hermite tensors, they are classified into two categories: those belonging to the basis (\ding{109}) and those which do not (\ding{55}). Lattice structures data are compiled from~\cite{SUCCI_Book_2002,SHAN_JCS_17_2016}.}
\label{tab:3DLattices}
\end{table}

Here the link between lattice structures and Hermite tensors is emphasized. To properly choose which Hermite polynomials should be taken into account in the expansion of $f_i$'s, the preservation of the orthogonality property of these polynomials, with respect to the weighted scalar product, is considered~\cite{PHILIPPI_PRE_73_2006}. As a reminder, this condition is as follows
\begin{equation}
\dfrac{1}{(2\pi rT_0)^{D/2}}\displaystyle{\int_{\mathbb{R}^D} \mathcal{H}^{(n)}_{\bm{\alpha}}\mathcal{H}^{(m)}_{\bm{\beta}} e^{-\left({\boldsymbol\xi^2}/{2rT_0}\right)}\,\text{d}\boldsymbol\xi} = \displaystyle{\sum_{i}\omega_{i}\mathcal{H}^{(n)}_{i,\bm{\alpha}}\mathcal{H}^{(m)}_{i,\bm{\beta}}} 
\end{equation}
with $c_s^2 = rT_0$. 
Results concerning some standard and high-order lattice structures are reported in Tabs.~\ref{tab:2DLattices} \&~\ref{tab:3DLattices}, for the 2D \& 3D cases respectively. The convention from~\cite{SHAN_JFM_550_2006} was used to describe the properties of each lattice structure: $E_{D,Q}^V$ where $D$ is the number of physical dimensions, $Q$ is the degree of precision of the quadrature, and $V$ is the number of discrete velocities. Furthermore, all velocities obtained by cyclic permutations and/or reflections with respect to each axis are omitted for the sake of clarity.\\
To conclude this appendix, expressions of the VDFs for the most complex lattice structures described herein ($E_{2,9}^{37}$ and $E_{3,9}^{103}$) are given by:
\begin{align*}
f_i^{(0),2D} = \omega_i \bigg[&\mathcal{H}^{(0)}_i a_{0}^{(0)} + \dfrac{1}{c_s^2}\left(\mathcal{H}^{(1)}_{i,x} a_{0,x}^{(1)} +  \mathcal{H}^{(1)}_{i,y} a_{0,y}^{(1)}\right)+ \dfrac{1}{2c_s^4}\left(\mathcal{H}^{(2)}_{i,xx} a_{0,xx}^{(2)} + 2 \mathcal{H}^{(2)}_{i,xy} a_{0,xy}^{(2)} + \mathcal{H}^{(2)}_{i,yy} a_{0,yy}^{(2)}\right) \\[0.1cm]
&+ \dfrac{1}{6c_s^6}\left(\mathcal{H}^{(3)}_{i,xxx} a_{0,xxx}^{(3)} + 3 \mathcal{H}^{(3)}_{i,xxy} a_{0,xxy}^{(3)} + 3 \mathcal{H}^{(3)}_{i,yyx} a_{0,yyx}^{(3)} + \mathcal{H}^{(3)}_{i,yyy} a_{0,yyy}^{(3)}\right)  \\[0.1cm]
&+ \dfrac{1}{24c_s^8}\left(\mathcal{H}^{(4)}_{i,xxxx} a_{0,xxxx}^{(4)} + 4 \mathcal{H}^{(4)}_{i,xxxy} a_{0,xxxy}^{(4)} + 6 \mathcal{H}^{(4)}_{i,xxyy} a_{0,xxyy}^{(4)} + 4 \mathcal{H}^{(4)}_{i,yyyx} a_{0,yyyx}^{(4)} + \mathcal{H}^{(4)}_{i,yyyy} a_{0,yyyy}^{(4)}\right) \bigg],
\end{align*}
and for the 3D extension,
\begin{align*}
f_i^{(0),3D} = f_i^{(0),2D} &+ \omega_i \bigg[\dfrac{1}{c_s^2}\mathcal{H}^{(1)}_{i,z} a_{0,z}^{(1)}+ \dfrac{1}{2c_s^4}\left( \mathcal{H}^{(2)}_{i,zz} a_{0,zz}^{(2)} + 2 \mathcal{H}^{(2)}_{i,xz} a_{0,xz}^{(2)} + 2 \mathcal{H}^{(2)}_{i,yz} a_{0,yz}^{(2)} \right) \\[0.1cm]
&+ \dfrac{1}{6c_s^6}\left(\mathcal{H}^{(3)}_{i,zzz} a_{0,zzz}^{(3)} + 3 \mathcal{H}^{(3)}_{i,zzx} a_{0,zzx}^{(3)} + 3 \mathcal{H}^{(3)}_{i,zzy} a_{0,zzy}^{(3)} + 3\mathcal{H}^{(3)}_{i,xxz} a_{0,xxz}^{(3)} + 3\mathcal{H}^{(3)}_{i,yyz} a_{0,yyz}^{(3)} + 6\mathcal{H}^{(3)}_{i,xyz} a_{0,xyz}^{(3)} \right)  \\[0.1cm]
&+ \dfrac{1}{24c_s^8}\Big(\mathcal{H}^{(4)}_{i,zzzz} a_{0,zzzz}^{(4)} + 4 \mathcal{H}^{(4)}_{i,zzzx} a_{0,zzzx}^{(4)} + 4 \mathcal{H}^{(4)}_{i,zzzy} a_{0,zzzy}^{(4)} +  6\mathcal{H}^{(4)}_{i,xxzz} a_{0,xxzz}^{(4)} + 6\mathcal{H}^{(4)}_{i,yyzz} a_{0,yyzz}^{(4)} \\[0.1cm]
&+ 12\mathcal{H}^{(4)}_{i,xxyz} a_{0,xxyz}^{(4)} + 12\mathcal{H}^{(4)}_{i,yyxz} a_{0,yyxz}^{(4)} + 12\mathcal{H}^{(4)}_{i,zzxy} a_{0,zzxy}^{(4)}\Big) \bigg].
\end{align*}
In the case of $f^{(1)}_i$ the very same terms, as for the equilibrium part  $f^{(0)}_i$, have to be taken into account. 

\twocolumngrid

\bibliography{COREIXAS_PRE_2017_Bibliography}

\end{document}